\documentclass[11pt]{article}         

\usepackage {epsfig,amsfonts,citesort}               

\begin{document}


\title{Null Cones in Lorentz-Covariant General Relativity\footnote{This work is based on a portion of the first's author's dissertation, supervised by the second author.} }
\author{J. Brian Pitts\footnote{The Ilya Prigogine Center for Studies in
     Statistical Mechanics and Complex Systems, 
     The University of Texas at Austin,
     Austin, TX 78712 USA,  and
	Mathematics Department, Box 780, 
	St. Edward's University,
	Austin, TX 78704 USA; 
	email jpitts@physics.utexas.edu or brianp@admin.stedwards.edu}
\footnote{J.B.P. thanks 
I. Goldman for providing results from his unpublished dissertation in Hebrew; M. Choptuik, L. Shepley, M. Visser, S. Deser, A. A. Logunov, S. Weinstein, J. Norton, and J. VanMeter  for discussions; C. V. Marcotte for the reference to the work of Weeks;  A. N. Petrov for bibliographic information and helpful discussions; and the Center for Relativity at the University of Texas at Austin for use of computer facilities. Support from the  Ilya Prigogine Center for Studies in Statistical Mechanics and Complex Systems at the University of Texas at Austin and from the Welch Foundation is very gratefully acknowledged.}
  and W. C. Schieve\footnote{The Ilya Prigogine Center for Studies in
     Statistical Mechanics and Complex Systems, 
     The University of Texas at Austin
     Austin, TX 78712 USA} }

\date{\today}

\maketitle


\begin{abstract}

     The oft-neglected issue of the causal structure in the flat spacetime approach to Einstein's theory of gravity is considered. Consistency requires that the flat metric's null cone be respected, but this does not happen automatically.  After reviewing the history of this problem, we introduce a generalized eigenvector formalism  to give a kinematic description of the relation between the two null cones, based on the Segr\'{e} classification of symmetric rank 2 tensors with respect to a Lorentzian metric. Then we propose a method to enforce special relativistic causality by using the naive gauge freedom to restrict the configuration space suitably.  A set of new variables just covers this smaller configuration space and respects the flat metric's null cone automatically.  In this smaller space, gauge transformations do not form a group, but only a groupoid.  Respecting the flat metric's null cone ensures that the spacetime is globally hyperbolic, indicating that  the Hawking black hole information loss  paradox  does not arise.
\end{abstract}





\section{Introduction}

     A number of authors have
discussed the utility of a flat
background metric $\eta_{\mu\nu}$ in general relativity or the possibility of deriving that theory, approximately or exactly, from a flat
spacetime 
theory \cite{Fierz,FierzPauli,Rosen1,RosenAnn,Tonnelat,TonnelatConference,Weyl,Papapetrou,PapapetrouUrich,Moshinsky,Kohler,Kohler2,GuptaPPSL,Gupta,GuptaReview,GuptaSupplement,Kraichnan,Kraichnan2,BelinfanteSwiss,BelinfanteSwihart,BelinfanteSwihart3,BelinfanteReview,BelinfanteCK,BelinfanteGarrison,ThirringFdP,Thirring,ThirringAPA,ThirringTrieste,Gutman,SexlConf,Sexl,NSSexlField,NSSexlLinear,Mandelstam62,Burlankov,Halpern,HalpernComp,KlauderCov,Huggins,Feynman,Feynman63,Feynman72,Arnowitt,Cornish,Weinberg64a,Weinberg64b,Weinberg64c,Weinberg64d,Weinberg64,Weinberg65,WeinbergTrieste,Weinberg,WeinbergWitten,PenroseWeinberg,Wyss,WyssStress,WyssGC,OP,OPMassless2,OPMassive2,Barbour,MittelstaedtCosm,Mittelstaedt,MittelstaedtDual,Westpfahl,Anderson,van Nieuwenhuizen,van N cones,Rovelli,Groenewold,DeserTrubatch,Deser,DeserTopology,DeserQG,Deser1,DeserConsist,DeserFermion,DeserCurved,DeserNepomechieLet,DeserNepomechie,DeserRosenFest,Deser3Half,PugachevGunko,Fronsdal,Cavalleri,Drummond,Penrose,Meszaros,Grishchuk,Zel1,Zel2,Grishchuk90,BurlankovCause,AvakianBimetric,LogunovFund,LogunovBasic,LogunovBook,PetrovNarlikar,Katz,KatzLerer,KatzBLB,PetrovKatz1,PetrovKatz2,Petrov,Petrov2,VlasovIncorrect,PetrovHarmonic,VlasovGaugeFix,LoskutovSchwarzschild,PopovaPetrov,GrishchukPetrov,PugachevGunko,Avakian,Gottlieb,Alvarez,Baryshev,Nikolic,NikolicTime,Nikishov,Babak,Straumann,SliBimGRG,Magnano,Scharf,Borne,Kiselev,BoulangerEsole}.  Some have permitted the background metric to be curved \cite{DeWitt67b,DeWittQGeom,DeWitt72,Barnebey,Roxburgh,Grishchuk,DeserCurved,KatzBLB,PetrovKatz1,PetrovKatz2,FatFerFra}, but our interest is in flat backgrounds only, because they are uniquely plausible as nondynamical entities. The use of a  background metric enables one to formulate a gravitational stress-energy tensor
\cite{Babak}, not merely a pseudotensor, so gravitational energy and momentum are localized 
in a coordinate-independent (but gauge-variant) way.  It also enables one to \emph{derive} general relativity and other generally covariant theories from plausible special-relativistic postulates, rather than
 postulating them \cite{GuptaPPSL,Gupta,GuptaReview,GuptaSupplement,Kraichnan,Kraichnan2,ThirringFdP,Thirring,ThirringAPA,ThirringTrieste,SexlConf,Sexl,NSSexlField,NSSexlLinear,Halpern,HalpernComp,Huggins,Feynman,Feynman63,Arnowitt,Weinberg64a,Weinberg64b,Weinberg64c,Weinberg64d,Weinberg64,Weinberg65,WeinbergTrieste,Weinberg,WeinbergWitten,PenroseWeinberg,Wyss,WyssStress,WyssGC,OP,OPMassless2,OPMassive2,Groenewold,Deser,DeserTopology,DeserQG,Deser1,DeserFermion,DeserCurved,DeserState,Fronsdal,Cavalleri,Meszaros,Grishchuk,Zel1,Zel2,Grishchuk90,LogunovFund,LogunovBasic,LogunovBook,Alvarez,Baryshev,Straumann,SliBimGRG,Magnano,Scharf,Borne,BoulangerEsole}. It is worth recalling a conclusion of E. R. Huggins \cite{Huggins}, who was a student of Feynman.  Huggins found that the
 requirement that energy be a spin-two field coupled to the stress-energy tensor does not lead to a unique theory, because of superpotential-type terms.  Rather, ``an additional restriction is necessary.  For Feynman
this restriction was that the equations of motion 
be obtained from an action principle; Einstein required that the gravitational field have a geometrical interpretation.  Feynman showed these two restrictions to be equivalent.'' \cite{Huggins} (p. 3)  
Because other derivations have built in the requirement of an action principle already, it is no surprise that Riemannian geometrical theories are the unique result.  As W. Thirring observed, it is not clear \emph{a priori} why
Riemannian geometry is to be preferred over all the other sorts of geometry that exist, so a derivation of effective Riemannian geometry is attractive \cite{Thirring}. 

	Casting  gravitation  in the same form as the other forces provides another reason to consider a field approach:  the flat metric's null cone provides a causal structure for defining dynamics in quantum gravity, which otherwise is lacking. 
The lack of an \emph{a priori} fixed  causal
structure is merely technically demanding at the classical level, but it constitutes a real puzzle at the quantum level, for one no longer knows how to write
equal-time commutation relations, for example, because one needs to know the metric in order to
determine equal times, but the metric is itself quantized:  a chicken-and-the-egg problem.

	Such derivations of general relativity and related theories, however, are perhaps only \emph{formally} special relativistic, because the curved null cone might not respect the flat one.  
This difficulty afflicts not merely our derivation \cite{SliBimGRG}, but in fact all derivations in this tradition, and implies that the alleged resemblance of Einstein's theory to other field theories in this approach is merely formal, for all that has been shown to date. 
We survey in some detail the 
treatment of this fundamental question over the last six decades.   As it happens, this issue has in general been ignored, explained away, postponed with the hope that it would go away, or mishandled,  although there have been positive signs in recent years.  We critique  claims that the problem is insoluble and  claims that it has already been solved, and conclude that the issue remains quite open.   

 Next we undertake to solve the problem.  The kinematic issue of the relationship between the two null cones is handled using the work of G. S. Hall and collaborators on the Segr\'{e} classification of symmetric rank 2 tensors with respect
 to a Lorentzian metric.  For our purposes, we classify the curved metric with respect to the flat one, and find necessary and sufficient conditions for a suitable relationship. Requiring that flat spacetime causality not be violated, and not be arbitrarily close to being violated, a condition that we
 call ``stable $\eta$-causality'', implies that all suitable curved metrics have a complete set of generalized eigenvectors with respect to the flat metric, and that the causality conditions take the form of \emph{strict} rather than loose inequalities. Given strict inequalities, one is in a position to solve such conditions, which are somewhat analogous to
 the ``positivity conditions'' of canonical gravity, which have been discussed by J. Klauder, F. Klotz, and J. Goldberg.  
In these new variables, stable $\eta$-causality holds \emph{identically}, because the configuration space has been reduced (though the dimension is unchanged), largely by reducing the lapse until the proper null cone relation holds.  This reduction implies the need for reconsidering the gauge freedom of the theory.  It turns out that gauge transformations no longer form a group, because multiplication is not defined between some elements.  But they do form a groupoid, which seems quite satisfactory.  Given the satisfactory outcome of the effort to make the proper null cone relationship hold, the above-mentioned derivations of general relativity as an ostensibly special relativistic theory in fact succeed.  The naive gauge freedom turns out to include some unphysical states, but that is not a serious problem.  
 
	Making the curved metric respect the flat null cone ensures that the resulting spacetime is \emph{globally hyperbolic}.  This fact is quite consistent with the existence of a region of no escape, which can arise due to the inward tilting of the curved null cones \cite{PetrovHarmonic}.  Given that global hyperbolicity apparently pulls the fangs from the Hawking black hole information loss paradox, it appears that this paradox does  not afflict the special relativistic approach to Einstein's theory of gravitation.  This lack of paradox indicates that the flat metric can be employed such that it is not merely a formal mathematical trick, but rather has beneficial physical consequences.


\section{Bimetric General Relativity and Null Cone Consistency: A History Since 1939}

     As we have seen, the use of a flat metric tensor $\eta_{\mu\nu}$ in gravitation has received a fair amount of attention over the last
 six decades or so.   However, the interpretation of the
resulting bimetric or field formulation of general relativity has not been adequately clarified, due to an ambiguous notion of causality:
 the effective curved metric which determines matter propagation is not obviously consistent with the flat background causal structure.  Having a consistent relationship is clearly a \emph{necessary} condition
 for a true special-relativistic theory. 

  Whether it is \emph{sufficient} is unclear from anything said so far, because the propagation of gravity itself or of nonminimally coupled matter fields can yield  more subtle behavior \cite{ClaytonCause,VisserVSL,Shore}.  However, we will see below that this null cone condition indeed is sufficient due to the presence of a well posed initial value formulation.

	 We now sketch the history of the flat null cone issue from roughly the late 1930s till the 
present.  We do not consider the period between 1905 and the late 1930s, though that might be an interesting 
project, which would consider the time after special relativity had solidified and include the invention of Einstein's theory of gravitation.  (L. P. Grishchuk mentions a bit of the history in works of Poincar\'{e} and 
Einstein \cite{Grishchuk90}.  Fang and Fronsdal sketch the history of the flat spacetime approach up to 1979 \cite{Fronsdal}, but neglect to consider the null cone issue.)  Rather, we start with the \emph{rebirth} of the flat spacetime approach to gravity, with works by Fierz, Pauli, and Rosen.
While the importance of the problem perhaps seems evident in retrospect, the neglect of it in the literature suggests that it in practice was not so obvious, or that influential radical empiricist philosophies obscured it.  One can roughly divide the issue's history into three periods, though at times we will disregard the historical boundaries to be able to discuss an author's whole work in a unified way. 
For the first 20 years (1939-1959), the problem seems not to have been recognized or mentioned in print (to our knowledge).  For the next two decades (1959-1979), it was sometimes mentioned, but either resolved incorrectly, dismissed as unimportant, or postponed with the hope 
that it would disappear. 
 More recently (1979-2001), it has been recognized more often, and occasionally regarded as worthy of sustained attention.  A few authors have attempted either to solve it or to prove it insoluble.  However, we disagree that either of these goals has been achieved, and will undertake to show why. One also finds authors continuing to
 ignore this problem even in the last few years \cite{WyssStress,Scharf}, and even within a review article \cite{Baryshev}.   When a fundamental issue is either ignored or mishandled for a long time, a critical history of the subject becomes necessary, so we provide it here.  

	One should perhaps distinguish between two null cone problems.  The first is:  given that one regards Einstein's equations as describing the evolution of an effective curved metric in Minkowski spacetime, what does one make of the potential violation of Minkowski causality by matter responding to the curved metric?  The second is: given that one chooses to quantize the geometrical (single-metric) theory, what does one make of causality without a metric to define equal-time commutation relations?  However, these problems are related, and we believe that the SRA as presented here solves them both, so we will treat them together.

\subsection{The Years 1939 to 1959: the Null Cone Consistency Problem Ignored}

	 Around 1940, in his seminal papers on the bimetric description of general relativity, N. Rosen  suggested that there ought to be some (gauge-fixing) relation between the flat and curved metrics,
 because one expects that the two coincide if the 
gravitational field vanishes \cite{Rosen1} (p. 149).  While this paper did not consider the meaning of the bimetric formalism in detail, its companion paper (p. 150) considered interpretive issues.  Rosen wrote (apart from a change in notation to match ours), \begin{quote} [f]rom the standpoint of the general
 theory of 
relativity, one must look upon $\eta_{\mu\nu}$ as a fiction introduced for mathematical convenience.  However, the question arises whether it may not be possible to adopt a different point of view, one in which the metric tensor $\eta_{\mu\nu}$ is given
 a real physical significance as describing the geometrical properties of space, which is therefore taken to be flat, whereas the tensor $g_{\mu\nu}$ is to be regarded as describing the gravitational field. \cite{Rosen1} (p. 150).\end{quote}
Rosen recognized that the flat spacetime view implies that the speed of light measured with ideal rods and clocks undistorted by gravity will differ from unity \cite{Rosen1} (p. 153), but he seems not to have addressed the possibility that it
 might \emph{exceed} $1$.  While his approach merely postulated bimetric general relativity, he did suggest that it would be desirable to derive it independently \cite{Rosen1} (p. 153).  His 
intention to carry out this procedure himself \cite{Rosen1} (p. 153) seems not to have been realized, but many others have done it since that time, as we have seen.

	 During the 1940s, with some war-time inconvenience in Greece, A. Papapetrou was able to express general relativity in an attractive form resembling electromagnetism, with the theories being expressed in the tensorial DeDonder and Lorentz gauges,
 respectively \cite{Papapetrou}.  He emphasized the improved nature of the conservation laws, especially for angular momentum, and found that certain attractive relations that have no invariant meaning in the geometrical view become perspicuous given 
the flat spacetime interpretation.  Papapetrou held that for the flat spacetime approach, gauge-fixing to tie together the two metrics was ``indispensable'' (p. 20), because the energy-momentum and angular momentum localization would suggest physically 
distinct systems given different relations between the two metrics.  He was aware of Rosen's result that the flat spacetime interpretation implies a varying speed of light (using unrenormalized instruments), but seems also to have failed to entertain the possibility 
that the gravitational field might make light travel \emph{faster} than in special relativity.   
	
	The neglect of the null cone issue continued well into the 1950s in the important works of S. N. Gupta \cite{GuptaPPSL,Gupta,GuptaReview,GuptaSupplement} and R. H. Kraichnan \cite{Kraichnan,Kraichnan2}.  At this stage the derivation of the exact nonlinearities of general relativity, which Rosen had desired, was achieved.   Concerning the special-relativistic nature of the 
theory, both authors seem to have regarded the Lorentz covariance of the theory as sufficient for special relativity.  If the theory's gauge invariance and the unobservability of the flat metric are mentioned, the idea that the observable effective curved metric might well \emph{conflict} with the flat metric is not.  This is an important distinction that will also be overlooked repeatedly by later authors.
One could imagine that the flat metric might fail to appear in the equations of motion, but still have its null cone serve as a bound on the curved metric's null cone, so this distinction is crucial.  The flat metric might have important qualitative consequences, even without having any quantitative role in the field equations.

	F. J. Belinfante, interested in the work of Papapetrou and Gupta and in particular in the solidifying covariant perturbation approach to quantizing gravity, contemplated the use of a flat metric in ``Einstein's curved universe'', which evidently
 meant the geometrical theory of gravity \cite{BelinfanteSwiss}.  Working in the context of the static Schwarzschild solution (in which it is difficult to get the relation between the two null cones wrong, at least outside the Schwarzschild radius, unless one tries to do so), Belinfante only had 
occasion to consider the null cone relationship incidentally.  But the fact that $r$ becomes $g$-temporal and $t$ becomes $g$-spatial for very small radii, juxtaposed with the \emph{a priori} fixed character of these quantities with respect to the flat metric, does give him pause.  Belinfante gives indications (including  in the paper's title) that he does not believe deeply in the flat spacetime approach, so perhaps the null cone 
issue would not have interested him.  While he is prepared to suggest that the ``Swiss-cheese''-like behavior of 
the Schwarzschild solution in the bimetric context might help eliminate field theory's divergences, it is clear from review papers on quantum gravity \cite{BelinfanteReview,BelinfanteCK} that the flat metric is just a tool--perhaps a useful one, but more likely not--for Belinfante. It is thus not too surprising that the null cone issue is ignored.
 The ``[r]eal problem''  is not to be found in ``[t]heories, usually in flat space, which seek to be approximations to Einstein's theory, or a perturbation-theoretical treatment 
of Einstein's theory'', but in ``[q]uantization of Einstein's theory itself.'' (pp. 198, 192) \cite{BelinfanteReview}.   For Belinfante, spacetime might have a Swiss cheese structure, contain worm holes, or have a closed spatial topology \cite{BelinfanteCK}.
Some of Belinfante's work with Swihart on linear gravity also neglects to discuss the null cone issue \cite{BelinfanteSwihart,BelinfanteSwihart3}.

\subsection{The Years 1959-1979:  the Problem Dismissed or Postponed}

	The null cone consistency issue is perhaps first discussed in print  by W. Thirring  in 1959 \cite{ThirringFdP,Thirring,ThirringAPA}, but then dismissed with a resolution that does not permit a true special relativistic interpretation.  Thirring clearly recognizes the apparent conflict between the two null 
cones  \cite{Thirring}, writing, ``Another feature of the equations of motion \ldots we want to point out is that the velocity $|d{\bf x}/dt|$ is not required to be $<1$ \ldots Thus [assuming the curved metric to be diagonal] 
there is a limiting velocity $c$ but it is space dependent and may exceed unity.'' (pp. 100, 101)  A bit later, he writes ``Since $c$ is also gauge dependent and will exceed unity in some gauge
 systems [the matter equation of motion] even admits an apparently acausal behavior.'' (p. 101)  However, Thirring thinks that this acausal behavior is \emph{only} apparent, for he is satisfied with the fact that the ``renormalized'' velocity (measured physically 
using real clocks, which are distorted by the gravitational field) is not greater than unity:  ``However, we shall see shortly that $c$ also corresponds to the velocity of light and that it becomes unity when measured with real measuring rods and clocks since they all are 
affected by the [gravitational] field.'' (p. 101)  Evidently the unobservable nature of the intervals governed by $\eta_{\mu\nu}$ satisfies Thirring that the apparently acausal behavior is not a problem: ``The real metric [interval corresponding to $g_{\mu\nu}$] is 
gauge invariant whereas [the interval corresponding to $\eta_{\mu\nu}$] is not and therefore has no physical significance.  Space-time measured with real objects will show a Riemannian structure whereas there are no measuring rods which could measure the original 
pseudoeuclidean space.'' (p. 103)  
Thirring's argument is doubtful because the same distinction that was neglected by Gupta and Kraichnan is also neglected here:  the non-measurability of the flat metric does not entail that it lacks physical significance.  Generally one considers causality to be
 an important 
  physical concept.  At the risk of stating the obvious, we recall that in special relativity, 
the relevant speed for causality is not the speed at which electromagnetic radiation actually propagates, 
 but the value of the universal velocity constant (ordinary called ``the speed of light'' and written as $c$, but to do so here would invite confusion) which appears in Lorentz transformations, that is essential.   
 As is well-known, to permit propagation faster than that speed in one frame is to admit backward causation--which is usually rejected--in another frame.  
Given the violation of the flat spacetime null cone, it is not clear what Thirring's field theoretic approach means.  Yet, according to  Thirring, the field theoretic approach gives
 ``a theory following the pattern of well understood field theories, in particular electrodynamics.'' (p. 116) 
 Thus, Thirring's list of advantages and disadvantages of the field and geometric approaches to gravity (pp. 116, 117) 
is notably incomplete, because the obvious notion of causality for the field approach has been discarded. Thirring comes very close to noticing the problem of null cone inconsistency, but then stops  short, apparently due to a prejudice against unobservable entities.

	One might hope that Thirring's almost-recognition of the problem would have inspired his successors to recognize and perhaps try to solve it.  That, however,
did not  occur.  In particular, although L. Halpern made a rather minute study of 
Thirring's paper \cite{HalpernComp}, the light cone issue receives only a single sentence (p. 388), one sufficiently noncommittal that no discomfort with Thirring's purported resolution of the causality issue is obvious.  Halpern was not an advocate of the flat spacetime approach to gravitation \cite{Halpern}, so it is the more remarkable that he overlooked a potentially  serious difficulty.  R. Sexl also was aware of the 
Thirring's work and even presented it at a conference \cite{SexlConf}, yet he also accepted Thirring's ostensible resolution  of the null cone conflict \cite{SexlConf,Sexl}.

	The covariant perturbation program for quantizing general relativity yielded a large number of works based on expanding the curved metric into a background part and a dynamical part.   Commonly the background metric 
was flat, leading to equations at least formally special relativistic.  Thus, one might expect the question of the relation of the two metrics to be considered in some way.  In fact, one finds rather less attention being given to this key conceptual issue than one would expect.

	A notable exception to the use of flat backgrounds is the work of B. S. DeWitt, who made great
 use of non-flat background metrics and found various benefits in doing so \cite{DeWitt67b}.  While DeWitt could make use of a background metric, to him it was always at most a tool, not a deep part of nature.  In an article entitled ``The Quantization of Geometry,'' he wrote:
\begin{quote} The problem of [quantizing the gravitational field] may be approached from either of two viewpoints, loosely described as the ``flat space-time approach'' and ``the geometrical approach.''  In the flat spacetime approach, which has been investigated by several
authors \ldots the gravitational field is regarded as just one of several known physical fields, describable within the Lorentz-invariant framework of a flat space-time.  Its couplings with other fields \ldots lead to a contraction or elongation of ``rigid'' rods 
and a retardation or advancement of ``standard'' clocks \ldots. Both the geometrical and flat space-time points of view have the same \emph{real} physical content.  However, it has been argued that the flat space-time approach provides more immediate access to 
the concepts of conventional quantum field theory and allows the techniques of the latter to be directly applied
to gravitation.  While there is merit in this argument, 
 too strong an insistence upon it would constitute a failure to have learned the lessons which special relativity itself has already taught.
Just as it is now universally recognized as inconvenient (although \emph{possible})
to regard the Lorentz-Fitzgerald contraction from relativistic modifications in the force law between atoms,
so it will almost certainly prove inconvenient at some stage to approach space-time geometry, even in the quantum domain, in terms of fluctuations of standard intervals which are the same for all physical devices and hence unobservable. \cite{DeWittQGeom} (pp. 267,268).\end{quote}
Concerning the question of a well-defined causal structure, which his approach appeared to lack, he later suggested, ``Critics of the program to quantize gravity frequency [\emph{sic}] ask, `What can this mean?' A good answer to this question does not yet exist.
However, there are some indications where the answer may lie.'' \cite{DeWitt67b}.  
DeWitt's vision for the program, which he was prepared to call ``covariant quantum geometrodynamics'' in a volume 
honoring J. A. Wheeler (the title itself suggesting sympathy for a geometrical view of gravitation, much as ``The Quantization of Geometry'' did), included that it ``should be able
 to handle any topology which may be imposed
 on $3$-space''\cite{DeWitt72} (p. 437).  Of the covariant perturbation formalism, he wrote that the ``most serious present defect of the covariant formalism 
is its foundation in scattering theory, with spacetime being assumed asymptotically flat.  The method of the
 background field, which we have introduced, indicates a way in which this defect
 may be removed'' \cite{DeWitt72} (p. 437). It seems very likely that the null cone issue, as we have formulated it, would not  be important to DeWitt, given that the background
 metric was merely a tool for investigating a truly geometrical theory. 

	Other authors, especially in the particle physics tradition, seem at least somewhat more content with a flat background metric.  In his lectures on gravitation, R. Feynman shows himself ambivalent about
 the interpretation of gravitation.  After deriving Einstein's equations from a flat spacetime field theory, he concluded from the unobservability of the flat metric that the latter was not essential.  Using an
 analogy with curiously intelligent insects walking on a tiled floor, he says that ``[t]here is no need to think of processes as occurring in a space which is truly Euclidean, since there is nothing physical which can ever be measured in this fictional space.  The tiles 
represent simply a labelling of coordinates, and any other labelling would have done just as well'' \cite{Feynman} (p. 101).  Concerning the ``assumption that space is truly flat,'' he concludes that ``[i]t may be convenient in order to write a theory in the beginning 
to assume that measurements are made in a space that is in principle Galilean, but after we get through predicting real effects, we see that the Galilean space has no significance'' (p. 112), but serves only as a ``bookkeeping device'' (p. 113).  Concerning 
the ``relations between different approaches to gravity theory,'' \begin{quote} [i]t is one of the peculiar aspects of the theory of gravitation, that it has both a field interpretation and a geometrical interpretation \ldots 
 these are truly two aspects of the same theory \ldots  the fact is that a spin-two field has this geometrical interpretation; this is not something readily explainable--it is just marvelous. 
 The geometric interpretation is not really necessary or essential to physics.  It might be that the whole coincidence might be understood as representing some kind of gauge invariance.  
It might be that the relationship between these two points of view about gravity might be transparent after we discuss a third point of view \ldots. ( p. 113) \end{quote}  Feynman seems to feel free to switch between the two views as he sees fit.  Questions about nontrivial topologies 
 or the desire to have a transparent notion of causality seem not to have occupied him.  Had they, he might have hesitated in proclaiming them to be ``the same theory'', given the competition between 
causality and gauge invariance.  Later, in developing the covariant perturbation theory, Feynman did not address these issues, but wrote as if no conceptual difficulties existed.  He wrote: ``The questions about making a `quantum theory of geometry' or
 other conceptual questions are all evaded by considering the gravitational field as just a spin-2 field nonlinearly coupled 
to matter and itself (one way, for example, is expanding $g_{\mu\nu} = \delta_{\mu\nu} + h_{\mu\nu}$ and considering $h_{\mu\nu}$ as the field variable) and attempting to 
quantize this by following
 the prescription of quantum field theory, as one expects to do with any other field.  The central difficulty springs from the fact that the Lagrangian is invariant under a gauge group,'' but this issue, he finds, can be resolved by adding a gauge-fixing term, 
the result being 
``completely satisfactory'' at the level of tree diagrams (which correspond to the classical theory) \cite{Feynman72}; see also (\cite{Feynman63}).  If the main difficulty is gauge invariance--which in fact competes with special 
relativistic causality, as Thirring nearly realized--and
 if the gauge-fixing terms lead to a ``completely satisfactory'' result without regard to the light cone relationship, then, unless we are to charge Feynman with oversight,  clearly the flat metric is merely a useful tool for him.  However, the claim to have avoided all conceptual questions cannot be sustained,
 because the light cone issue is just such a question,
and the meaning of parts of Lorentz-covariant field theory remains obscure if the problem is ignored.  Huggins, a student of Feynman, also neglects to consider the null cones issue \cite{Huggins}.

	 S. Mandelstam presented a critique of the flat-space covariant perturbation program as Gupta had developed it 
 \cite{Mandelstam62}.  Gupta had imposed the DeDonder coordinate (gauge) condition.  Let us see how close Mandelstam comes to identifying the null cones issue.  He writes:
``Quantization in flat space can only be regarded as a provisional solution of the problem for several reasons,'' such as its approximate (at least at that stage of development) character, the use of an indefinite metric, and the presence of unphysical states.  ``But the main objection to this method of quantization lies surely in the physical sacrifices it makes by going to flat space.  The variable specifying the coordinates are numbers without physical significance which can be chosen 
in an infinite variety of ways.  On the other hand, distances in space-time,  which are physically significant entities, are related to the coordinates in a manner which has
not been elucidated when the metric is quantized.''  However, perhaps these objections can be met:
``It may be possible to add to the theory a prescription for interpreting its results physically.  If it could then be shown that the predictions of the theory were independent of the coordinate 
conditions used, and that they tended to the predictions of the unquantized theory in the classical limit, we would have a satisfactory theory.  Some progress has actually been made in this direction by Thirring'', which ``indicates the connection of the Gupta variables to the metric,'' though ``the basic difficulties of the `flat space' approach remain.''  Clearly one of Mandelstam's worries is
the question of gauge invariance in a procedure that
makes use of coordinate conditions.  It is difficult to tease out a clear statement of worry about rival null cones from these remarks, though the issue might  have been intended among the ``the basic difficulties of the `flat space' approach'' that remain. 

	A moment of  clarity occurred in 1962 with the appearance of a paper by J. R. Klauder \cite{KlauderCov}, whose abstract opens with the statement, ``[i]n any quantum theory, in which the metric tensor of Einstein's gravitational theory is also quantized, it becomes meaningless to ask for an initial space-like surface on which to specify the conventional field commutators.''  Klauder elaborates:
\begin{quote}In so far as [certain] formalisms [for quantizing gravity] are transcriptions of techniques successful in a flat Lorentz space-time, they ignore a unique problem peculiar to general relativity.  Conventional field theories deal, in particular, with commutation rules, which, when employed for the fields separated by a space-like interval, have an especially simple form.  Whether two nearby points are
 or are not space-like is a \emph{metric} question that can be asked (and in principle answered) not only in flat space but also in any space with a preassigned curved metric as well.  However as soon as the space-time metric $g_{\mu\nu}(x)$ becomes a dynamical variable--as in Einstein's theory--then an initial space-like surface on which to specify commutators of any two fields becomes a meaningless concept.  
\end{quote}
Klauder's approach to handling this problem was to propose an alternative formalism in which fields  can fail to commute at most only at the same \emph{event}--a radical move in the opposite direction.  Unfortunately, Klauder's acute awareness of the null cone issue did not spread widely.

	S. Weinberg did considerable work on gravitation considered as a Lorentz-invariant theory \cite{Weinberg64a,Weinberg64b,Weinberg64c,Weinberg64d,Weinberg64,Weinberg65,WeinbergTrieste,Weinberg,WeinbergWitten}.  Concerning the geometric interpretation of general relativity, Weinberg could write that
  ``the geometric interpretation of the theory of gravitation has dwindled to a mere analogy, which lingers in our language \ldots but is not otherwise very useful.  The important thing is to be able to make 
predictions about images on photographic plates, frequencies of spectral lines, and so on, and it simply doesn't matter whether we ascribe these predictions to the physical effect of gravitational
 fields or to a curvature of space and time.'' \cite{Weinberg} (p. 147)  This ambivalence about the meaning of the theory perhaps helps to explain why the null cone consistency issue
appears to be ignored in Weinberg's writings.  However, the meaning of concepts used in Lorentz-invariant field theory in which Weinberg's work is rooted, or at least its relation to an underlying classical theory, does seem somewhat obscure if this issue is neglected.  Somewhat more recently, R. Penrose reported that Weinberg was ``no longer convinced that the 
anti-geometrical viewpoint is necessarily the most fruitful'' \cite{PenroseWeinberg}, on account of some impossibility theorems \cite{WeinbergWitten}.  In recent personal communication with one of us (J. B. P.), Weinberg stated that he is
 no longer a strong advocate of any view on the subject, though it is quite interesting that the flat spacetime approach reproduces general relativity.

	Based on the ``spin limitation principle,'' which requires that only definite angular momenta be exchanged, V. I. Ogievetsky and  I. V. Polubarinov have derived Einstein's 
equations and a family of massive relatives thereof in flat spacetime \cite{OP,OPMassless2,OPMassive2}.  While this principle is quite attractive, it fails to pay any heed to whether the resulting theories yield propagation consistent with the causal structure
 of the flat metric.  Given that some of their theories are massive and thus make the $\eta_{\mu\nu}$ \emph{observable}, this shortcoming seems fairly serious.  While we can find no mention of the null cone issue in the work of Ogievetksy and Polubarinov, it would be interesting to see if the spin limitation principle could be generalized in such a way as to yield consistency of the null cones.

	A large amount of work related to the field approach has been done by  S. Deser, sometimes with collaborators such as D. G. Boulware, R. Nepomechie, A. Waldron or others.  In the course of papers which
 derived general 
relativity via self-interaction in flat spacetime \cite{Deser} or curved \cite{DeserCurved}, or general relativity from quantum gravity \cite{DeserQG,Deser1}, or supergravity from self-interaction \cite{DeserFermion}, or which study bimetric theories 
for a festschrift for N. Rosen \cite{DeserRosenFest}, we can find no mention of the issue of the null cone consistency issue.  In particular, Deser finds the  main issues for  bimetric theories to be essentially the same problems that he and Boulware found in massive variants of general
 relativity \cite{DeserMass}, \emph{viz.}, empirically falsified light bending properties, negative energy disasters, or both \cite{DeserRosenFest}.  Unlike some authors who have a strong preference, Deser (after a rather pro-geometrical paper early on \cite{Deser57}) seems to admire both the geometric 
and field formulations: ``The beautiful geometrical significance of general relativity is complemented by its alternate formulation as the unique consistent self-coupled theory arising from flat-space free gravitons, without appeal to general covariance.'' \cite{DeserCurved} 
 However, depending on how one reads the flat spacetime approach, one might obtain some different features, as we will 
observe below, so one might prefer to see the meaning of the field formalism addressed.  In fact the closely related issue of the topology of spacetime does receive some attention \cite{DeserTopology}.  Clearly Deser is not interested in taking the flat background with the utmost seriousness, for he states that solutions to Einstein's equations not connected to Minkowski spacetime ought to be considered.  He even suggests that one need not allow for nontrivial topologies by hand, because possibly they arise automatically.  The question of respecting the conformal structure of the flat background does not arise.  Perhaps the closest one finds to discussion of the conformal structure is in the work by Deser and R. Nepomechie on the anomalous propagation of gauge fields in some conformally flat spacetimes, compared to a flat background \cite{DeserNepomechieLet,DeserNepomechie} with the same null cone structure.  In particular, backscattering 
off the geometry causes the propagation to lie not merely on the null cone, but inside it.  However, ``while our results are surprising, they do not imply any consistency problems'' \cite{DeserNepomechieLet},
 as they would if the propagation were \emph{outside} the null cone.\footnote{We thank Prof. Deser for calling our attention to this issue and the spin $\frac{3}{2}$ field.}  

	Some time ago it was recognized by G. Velo and D. Zwanziger  that spin $\frac{3}{2}$ field propagation has causality worries. They found that the ``main lesson to be drawn from our analysis is that special relativity is not automatically satisfied by writing equations that transform covariantly.  In addition, the solutions must not propagate faster than light'' \cite{VeloZwanziger}.  A rather similar lesson needs to be learned regarding the gravitational field (spin $2$) as well, for this is just the point that Gupta and Kraichnan missed.

	During the 1970s, the relation between the two null cones continued to be neglected in the context of the covariant perturbation approach to quantizing general relativity, at least in practice. 
 However, quantum gravity review talks drew attention to this problem from time to time.  
This service was performed with special clarity in a 1973 review by A. Ashtekar and R. Geroch \cite{AshtekarGeroch}.  They find that much of the difficulty in quantizing the theory arises from the fact 
that ``the distinction between the arena and the phenomenon, characteristic of other physical theories, is simply not available in general relativity: the metric plays both roles.'' \cite{AshtekarGeroch} (p. 1214)  In discussing field theoretic approaches, they write that \begin{quote}
[i]t is normally the case in quantum field theory  \ldots that two distinct fields come into play--a kinematical background field (the metric of Minkowski space) and a dynamical field \ldots. One can certainly regard general relativity as a field theory, but in
 this case there is only a single field, the metric $g_{ab}$ of spacetime, which must play both these roles.  But the application of the techniques of quantum field theory apparently requires a non-dynamical background field.  In quantum electrodynamics, for example, the 
causality of the Feynman propagators and the asymptotic states, in terms of which the $S$-matrix is defined, refer directly to the metric of Minkowski space.  Thus, one does not expect to be able to carry over directly to general relativity, regarded as a classical field
 theory, the procedure which led for example from classical Maxwell theory to quantum electrodynamics.  In order to apply the techniques of quantum field theory one must, apparently, either modify these 
techniques or reformulate the interpretation of general relativity as a field theory. (p. 1229) \end{quote}
This latter suggestion of reinterpreting general relativity does not strike us as being unimaginable or even excessively difficult, especially given that Kraichnan had presented a simple and clean derivation  already in 1955 \cite{Kraichnan}. 
However, Ashtekar and Geroch do present some objections to this general line of attack.  ``It turns out, however, that this perturbation approach to obtaining a quantum theory of the gravitational field suffers from a number of difficulties.  There exist, \cite{GerochSpinor} 
for example,
four-dimensional manifolds $M$ on which there are metrics $g_{ab}$ of Lorentz signature, but on which there are no flat metrics.'' (p. 1232)  However, it is not clear why such examples must be regarded as physically admissible.  If it could be shown that some exact solutions of obvious physical utility admit no flat metric, 
then the argument would be persuasive, but that argument was not made. Thus, it seems that this argument against the perturbation approach will be highly persuasive only if one is \emph{already} committed to a geometrical view of Einstein's equations at the classical level.  But one would exaggerate only slightly to say that
this is the point at issue.
The idea of \emph{requiring} that the curved null cone be consistent with the flat one seems not to have been entertained, but Ashtekar and Geroch have shown powerfully, if reluctantly, why such an approach merits consideration.

	The null cone consistency issue was emphasized  by C. J. Isham at the first Oxford quantum gravity symposium \cite{Isham1}.  We quote from pp. 20, 21: 
\begin{quote} One natural approach perhaps is to separate out the Minkowski metric $\eta_{\mu\nu}$ and write $g_{\mu\nu}(x) = \eta_{\mu\nu} + h_{\mu\nu}(x)$ where $h_{\mu\nu}(x)$  describes the deviation of the geometry from flatness \ldots [which approach has some advantages.]
However, there are a number of objections to this point of view.  For example: (i) The actual background manifold may not be remotely Minkowskian in either its topological or metrical properties, in which case the separation [above] \ldots is completely inappropriate.
  (ii)  Even if [the equation above] is justified (from the point of view of i)) [\emph{sic}] the procedure is still dubious because the lightcone structure of the physical spacetime is different from that of Minkowski space.  For example, if the field 
$\hat{\phi}$
has some sort of microcausality property with respect to the metric $g_{\mu\nu}$ then this is \underline{not} equivalent to microcausality with respect to the fictitious Minkowski background.\end{quote}
Once again a prior commitment to a geometrical view of general relativity is manifest.
Isham seems not to entertain the idea of \emph{requiring} that the spacetime be compatible with the flat background, but at least the consistency issue is clearly stated.
At the second Oxford symposium, Isham observed that ``one of the ambitions of the Riemannian programme is to free quantum gravity from perturbation theory based on the
 expansion $g_{\mu\nu} = g_{\mu\nu}^{c} + \sqrt{G} h_{\mu\nu}$.  Expansions of this type are known to be bad in classical general relativity and they clearly misrepresent the global topological and lightcone structures of the pair $(M, g_{\mu\nu})$'' \cite{Isham2} (p. 14). We suggest that no misrepresentation occurs if one chooses $(M, g_{\mu\nu})$ suitably.

 Isham has continued to mention this issue of null cone troubles in more recent talks in the context of the problem of causality and time \cite{IshamStructure,ButterfieldIsham}.  The problem of time shows up in the light cone issue for the covariant perturbation approach to quantum gravity, but related difficulties show up elsewhere \cite{IshamStructure}. 
 Still more recently, Isham has expressed the issue as follows:  \begin{quote}
	\emph{The problem of time}  The background metric $\eta$ provides a fixed causal structure with the associated family of Lorentzian inertial frames.  Thus, at this level, there is no problem of time.  The causal structure also allows a notion of microcausality, thereby permitting a conventional type of relativistic quantum field theory to be applied to the field $h_{\alpha\beta}$.

	However, many people object strongly to an expansion [of the curved metric into a flat one plus a dynamical part] since it is unclear how this background causal structure is to be related to the physical one; or, indeed, what the latter really means
 \ldots it is not clear what happens to the microcausal commutativity conditions in such circumstances; or, indeed, what is meant in general by `causality' and `time' in a system whose light cones are themselves the subject of quantum fluctuations.\cite{ButterfieldIsham} (p. 58) \end{quote}  While these are interesting questions, there seems to be no reason to think them unanswerable; indeed we shall substantially answer some of them.

	Another moment of awareness of the null cone consistency issue in a  quantum gravity review talk comes from P. van Nieuwenhuizen at the first Marcel Grossmann meeting.  After showing keen awareness of the problem, van Nieuwenhuizen shelves it.  He writes:
\begin{quote}
According to the particle physics approach, gravitons are treated on exactly the same basis
as other
particles such as photons and electrons.  In particular, particles
(including gravitons) are always in flat Minkowski space and move
\underline{as if} they followed their geodesics in curved spacetime
because of the dynamics of multiple
graviton exchange.  This particle physics approach is entirely equivalent to the usual geometric approach. Pure relativists often become somewhat uneasy at this point because
of the following two aspects entirely peculiar to gravitation:
\begin{itemize}
\item In canonical quantization one must decide before quantization
which points are
spacelike separated and which are timelike separated, in order to define the basic commutation relations.  However, it is only after quantization
that the
fully quantized metric field can tell us this spacetime structure.  It follows that the concept of space-like or time-like separation has to be preserved under quantization, and it is not 
 clear whether this is the case.  (One might wonder whether the causal structure of spacetime need be the same in covariant quantization as in canonical quantization.)
\item  Suppose one wanted to quantize the fluctuations (for example of a scalar field, or even of the gravitational field itself)
 about a given curved classical background instead of about flat Minkowski spacetime.  In order to write the field operators corresponding to these fluctuations in second-quantized form, one needs positive and negative frequency (annihilation and creation) solutions. 
 In non-stationary spacetimes it is not clear whether one can define such solutions.  (It may help to think of non-stationary space-time as giving rise to a time-dependent Hamiltonian.)
\end{itemize}  The strategy of
particle physicists has been to ignore these two problems for the time being, in the hope that they will ultimately be resolved in the final theory.  Consequently we will not discuss them any further.\cite{van N cones}
\end{quote}
He raised the issue again in another work \cite{van N renorm}.  While quantization is not our immediate concern, a similar worry to the first of these two exists at the  classical level if one wishes to take the flat metric seriously: there is no
 reason to expect  that the dynamics will  yield automatically a physical causal structure
consistent with the \emph{a priori} special-relativistic one, but inconsistency leads to grave interpretive difficulties.  As this history shows, workers in the covariant perturbation quantum gravity program gave most of their attention to technical issues, not conceptual questions (see also (\cite{Wallace}).

	
\subsection{The Years 1979-2001: the Problem Increasingly Attended  and the Development of Three Views}

	More recently, the question of null cone consistency has come to be recognized as interesting somewhat more often.  While a fair number continue to neglect the issue, those who have addressed it can be
 found to have one of three attitudes toward the flat metric: that it is  a useful fiction, that it is  a useless fiction, or that it is  the truth.  These views will be considered in turn.  First we note some recent signs of the growing
 awareness of the problem.  
	
	In the 1984, the subject made its way into a standard text \cite{Wald}.  R. Wald writes:  ``The breakup of the metric into a background metric which is treated classically and a dynamical field $\gamma_{ab}$, which is quantized, is unnatural from 
the viewpoint of classical general relativity.  Furthermore, the perturbation theory one obtains from this approach will, in each order, satisfy causality conditions with respect to the 
background metric $\eta_{ab}$ rather than the true metric $g_{ab}$.  Although the summed series (if it were to converge) still could satisfy appropriate causality conditions, the covariant 
perturbation approach would provide a very awkward way of displaying the role of the spacetime metric in causal structure.'' \cite{Wald} (p. 384). Once again, a prior commitment to a geometrical understanding of classical gravity is evident. Some of Wald's negative attitude toward
 the ``breakup'' of $g_{ab}$ results from assuming that the curved metric is fundamental, not derived.  But given how easy it is to \emph{derive} Einstein's equations from a flat spacetime theory \cite{Kraichnan,SliBimGRG}, why should one not regard the \emph{curved} metric as derived?  Be that as it may, one is pleased that the light cone issue is emerging from the neglect that it once suffered.  It is intriguing that Wald suggests that  the whole series might be $g_{ab}$-causal even 
though each term is $\eta_{ab}$-causal.  The most obvious way for such to occur 
 would be for the curved metric's null cone in fact to be confined on or 
within the flat one's.  If that is the case, then it seems that Wald is almost suggesting (albeit reluctantly) what we will do below.

	The recent contemplation of ``naive quantum gravity'' by S. Weinstein also has called attention to the lack of a fixed causal structure in quantum gravity \cite{Weinstein}.  If one is interested 
in full quantum gravity, as opposed to semiclassical work,  then ``we would expect that the metric itself is subject to quantum fluctuations \ldots But if the metric is [subject to quantum fluctuations], then it is by no means clear that it will be meaningful to talk about whether $x$ and $y$ are
 spacelike separated, unless the metric fluctuations somehow leave the causal (i.e. conformal) structure alone.'' (pp. 96-7) There appear to be two things that this last suggestion might mean.  First, it might mean that the metric is conformally flat, so that the causal structure is just that of flat spacetime, while gravity is described 
by a scalar field. However, it is well-known that scalar gravity is empirically falsified by the classical tests of general relativity \cite{Kraichnan}, so this must not be Weinstein's intent.  Second, it might mean that,
 although the full metric is allowed to vary, its variations are \emph{bounded} so that the null cone of the nondynamical (and presumably flat) metric is respected.  {Correspondence between J. B. P. and Weinstein makes clear that this in fact was intended.}  That is what we propose here. 
Weinstein does indeed consider ``whether it is at all \emph{possible} to construe gravitation as a universal interaction that nonetheless propagates in  flat, Minkowski spacetime.'' (p. 91)  He concludes that \begin{quote}
the short answer is, `No,' for three   
reasons.  First, the `invisibility' of the flat spacetime means that there is no privileged way to decompose a given curved spacetime into a flat background and a curved perturbation about that background. Though this non-uniqueness is not particularly problematical 
for the classical theory, it is quite problematical for the quantum theory, because different ways of decomposing the geometry (and thus retrieving a flat background geometry) yield different quantum theories.  Second, 
not all 
topologies admit a flat metric, and therefore
spacetimes formulated on such topologies do not admit a decomposition
into flat metric and curved perturbation.  Third, it is not clear \emph{a priori} that, in seeking to make a decomposition into background and perturbations about the background, the background should be \emph{flat}.  For example, why not use a background of constant curvature? (p. 92)
\end{quote} However, these arguments are less than compelling.  Concerning the first argument, Weinstein provides neither argument nor citation.  It appears to be a claim that a suitably gauge-invariant theory cannot be constructed.  But why should one believe that?  Concerning the second objection, which resembles that of Ashtekar and Geroch, the advocate of flat spacetime will ask ``why are nontrivial topologies necessary?''  There are no \emph{facts} or even good arguments that require them at present.\cite{Meszaros}  In the absence of such, the insistence that nontrivial topologies are  theoretically necessary is close to question-begging.  For why not merely adopt a nongeometrical view of gravity at the classical level, too?\footnote{The issue of the topology of the universe has seen a fair amount of attention lately, with the aim of experimental test--see, for example,  \cite{Weeks1,Weeks2,Blanloeil,ReboucasTop,Roukema,Levin}.}   Concerning the last objection, it seems clear to us that a flat background is the default choice because it is simpler than any
other choice.    While any other choice requires some argument for making that choice instead of the others and strongly suggests the question ``why does spacetime have \emph{this} geometry?'', flat spacetime does not, but rather is the obvious default choice. Weinstein's specific alternative suggestion 
of a constant curvature spacetime, for example, suggests the question ``why does the curvature take \emph{this} value, as opposed to some other value?'' There is a one-parameter family of constant curvatures that one might specify, but flat spacetime, requiring no parameter, is simpler.  We can agree with Weinstein that in ``allowing metric fluctuations to affect causal structure, one is clearly at some remove
 from ordinary field-theoretic quantization schemes.'' (p. 97)  But it seems unclear, \emph{pace} Weinstein, that there is any need to renounce the use of a flat background causal structure.  

	It is  unfortunate that some recent articles  still do not address the null cone issue.  However, enough have done so that one can identify three major attitudes toward the use of a flat metric that one finds.  One view is that the flat metric is a useful fiction.  Another, more purely geometrical view holds that the flat metric is a useless fiction.  The third view regards the flat metric as the truth.  We survey these approaches in this order.

\subsection{Field Formulation: the Flat Metric as a Useful Fiction}

	Some authors have explicitly stated that the flat metric is merely an auxiliary object, formally useful but not tied to the causal structure of the theory \cite{Zel1,Zel2,Grishchuk90,PetrovNarlikar,Petrov}.  The reasons given include the gauge-variance of the 
relationship between the null cones and the unobservability of the flat metric.  The first  objection will be answered below by a new definition of a gauge transformation.  These authors seem to have assumed that gauge transformations should form a group, whereas we find a groupoid to be a natural substitute.  The fact that the flat metric's 
null cone is sometimes violated using otherwise-convenient gauges appears to be another reason: it does not appear possible to fix the gauge to be, say, tensorial DeDonder and have the null cone relationship be automatically satisfactory.  L. P. Grishchuk has written that ``the mutual disposition of the light cones of the $g_{\mu\nu}$ 
and $\eta_{\mu\nu}$ can be of interest only in the case when the attempt is made to interpret the metric relations of the world as observable,'' which efforts are of course bound to fail, he continues \cite{Grishchuk90}.  Unfortunately, Grishchuk 
has overlooked the same distinction that Gupta, Kraichnan, Thirring, Feynman, and probably many others missed,  and has failed to recognize that if the null cones can be made consistent, then a conceptual difficulty posed by quantization would be eliminated. 
 
	A. N. Petrov describes the same view (though we have taken the liberty of spelling out with words the abbreviations
 used): \begin{quote} However, the background in the field formulation of general relativity is not observed.  The movement of test particles and light rays is not connected with the geometry of the background 
spacetime.  The light velocity in the
 background spacetime can approach an infinite value.  In contrast, in the geometrical formulation of general relativity the test particles and the light rays 
define the geodesics in real physical spacetime.  Thus, the background spacetime in the field formulation of general relativity is an auxiliary and nonphysical (fictitious) concept \ldots  We stress that the field formulation of general relativity and the geometrical formulation of general relativity are two different formalisms for a description of the same
 physical reality and they lead to the same
 physical conclusions \ldots there are no obstacles in treating any solution to general relativity (spacetime) in the framework of the field formulation
 of general relativity.  However, it is 
clear that a manifold which supports a physical metric will not coincide in general with a ``manifold'' which supports an auxiliary metric.  As a result, in the field configuration on the auxiliary nonphysical background, ``singularities,''
 ``membranes,'' ``absolute voids,'' and others can appear.  This leads to cumbersome and confused interpretations and explanations.  Thus, the whole spirit of general relativity itself requires 
 the investigation of many problems with the help of the geometrical formulation technique.  However, there exists problems [sic] for an investigation in which the field formulation 
technique is more convenient \cite{Petrov} (pp. 452, 453).\end{quote}
  Thus, the field formulation is seen as a tool that sometimes is helpful,  but sometimes not so convenient, and in any case not to be trusted in addressing deep issues. 

	A similar attitude has been taken by D. E. Burlankov \cite{BurlankovCause}, who did some early work using a flat background metric as a convenient fiction \cite{Burlankov}.  Burlankov objects to the fundamental status of Minkowski spacetime because of
 the gauge-variance of the null cone relation, and also because the curved null cone \emph{differs} from the flat null cone \cite{BurlankovCause}.  The former argument will be addressed in due time.  The latter argument, in Burlankov's hands, is said to  imply that only 
 curved metrics conformally related to the flat background would be acceptable.  But this objection is just unpersuasive.  

\subsection{Geometrical Formulation:  the Flat Metric as a Useless Fiction}

	Other authors have taken the view that the flat metric is a blemish on the pure geometric beauty of general relativity, and thus is to be avoided.  Such a description would seem to 
fit R. Penrose \cite{Penrose} and J. Bi\v{c}\'{a}k \cite{Bicak}.   This negative attitude toward the flat metric seems to have motivated Penrose to note that the null cone issue really must be handled if the Lorentz-covariant approach (which he associates with Weinberg) is to be considered 
satisfactory.  Penrose, recognizing 
the connection between scattering theory and the Lorentz-covariant perturbation approach to gravity,  poses a dilemma for the  latter. 
 Using global techniques, he shows that either the curved null cone locally violates the flat one, or the scattering properties become inconvenient because the geodesics for the two metrics
 continue to diverge even far away from a localized source.  He concludes 
that a ``satisfactory'' relationship between the two null cones cannot be found.  
 Concerning the horns of Penrose's dilemma, we simply accept the second one.  It is known that long-range fields have inconvenient scattering properties \cite{Goldstein}.  
We find that the root of the divergence between the geodesics is 
merely the long-range $\frac{1}{r}$ character of the potential in the conformally invariant part
of the curved metric.  If the fall-off were a power law of the form $\frac{1}{r^{1 + \epsilon} }$, $\epsilon > 0$, then no difficulty would arise.  
So this objection is basically a reflection of the fact that a long-range symmetric tensor potential exists.    But why is that a fundamental problem?  

\subsection{Special Relativistic Approach:  the Flat Metric as the Truth} 

	Besides the field-in-fictitious flat spacetime and
 geometrical approaches, there is another attitude that one might take toward the flat metric approach, \emph{viz.}, that special relativity is correct in its usual strict sense 
(global Lorentz invariance, trivial spacetime topology, and no violation of $\eta$-causality), and thus that the gravitational field must be made to respect the flat metric's causal structure. 
 This view is more conservative than the other views \cite{Feynman} (p. 101), and is sufficiently obvious and attractive an idea that one might  expect it to have been explored thoroughly,
 probably decades ago, and either sorted out or refuted.  But that expectation would be disappointed.  Demonstrating this surprising fact was one purpose of the substantial review of the history of the subject above.   Some authors have claimed to have sorted it out, and some to have refuted it, but we disagree on both points.  

	For the sake of convenience, this approach needs a name.  We will use the term ``special relativistic approach'' (SRA).  We have resisted calling this approach a ``formulation'' to match Petrov's ``field formulation'' and ``geometric formulation,'' because it turns out that
the SRA is in fact physically distinct, though in rather subtle and recondite ways, from the geometrical approach.  The SRA takes a realist attitude toward the flat metric, whereas the field formulation takes an instrumentalist attitude.
Some have objected to
regarding a 
theory based on the Einstein equations 
as something other than general relativity \cite{Zel1,Zel2,Grishchuk90}.   Others, including J. Norton, have insisted
 that the SRA is distinct from general relativity.
 Perhaps the common usage of the term ``general relativity'' is simply 
too vague to provide a 
 resolution to this difficulty.  If nothing else were at stake, one would avoid pretentious claims of a new theory.  But as will appear below, there is in fact be a physical difference, which might even be testable in exotic circumstances, 
between the two approaches.  The phenomena of collapse to form black holes seem to be altered somewhat in the SRA, not least because the SRA implies global hyperbolicity.

	Let us recall some common terminology regarding lengths and times given by the two metrics.  Those measured using ordinary rods and clocks will be distorted by gravity and thus be governed by $g_{\mu\nu}$.  These are ``renormalized'' measurements.  Lengths and times `measured' using ideal rods and clocks, which are not distorted by gravity, will be governed by $\eta_{\mu\nu}$.  These are ``unrenormalized'' measurements.  Let us also write  $ds^{2}$ for the renormalized $g$-interval and $d\sigma^{2}$
 for the unrenormalized $\eta$-interval.  These two types of lengths and times have different uses.  Renormalized measurements will serve for all mundane purposes.  Unrenormalized measurements, on the other hand, will be relevant in more mathematical or metaphysical contexts.  For example, in deciding whether a solution of Einstein's equations exactly covers Minkowski spacetime, one must unrenormalized measurements.  The solution should be neither too small for Minkowski spacetime, nor too large for it.  Flat Robertson-Walker models with finite renormalized age appear too small, because naively it appears that the early end of Minkowski spacetime just lacks a curved metric.  This problem can likely be resolved by stretching the Robertson-Walker solution so that the singularity occurs at unrenormalized past infinity.  This move resembles one by  C. Misner \cite{MisnerTime}, A. G. Agnese and A. Wataghin \cite{Agnese}, and J.-M.  Levy-Leblond \cite{Levy-Leblond}, but our introduction of a flat background metric gives a compelling physical motivation, which they perhaps lacked.  

	On the other hand, exact gravitational plane wave solutions appear to be too large for Minkowski spacetime.  Whereas monochromatic linearized plane wave solutions in the Hilbert gauge give local causality violation, exact plane wave solutions can yield a global violation of causality due to lack of a Cauchy surface.  These solutions have been called ``caustic plane  waves'', and were first noted by Penrose \cite{PenrosePlane}. Bondi and Pirani write, \begin{quote}
We call a plane gravitational wave caustic if it is capable of inducing accelerations of test particles so strong that two particles, initially at rest in flat spacetime and aligned suitably with respect to the wave, but arbitrarily far apart, will collide, within
 a finite time interval that is independent of their initial separation, after being struck by the wave. We shall show that all plane sandwich waves with fixed polarization are caustic, and describe some unusual optical properties of such waves \ldots. \cite{BondiCaustic} (p. 395).\end{quote}  We recall that a sandwich wave is one having a limited duration, so the wave  is `sandwiched' between two regions of flat spacetime.
A special case of this caustic phenomenon is found in W. Rindler's text \cite{Rindler} (pp. 166-174).  Such a phenomenon is obviously inconsistent with the SRA.  In a set of coordinates that one rather naturally chooses, there exists a coordinate singularity in this solution \cite{Rindler}. In 1937  N. Rosen concluded that plane waves in general relativity do not exist, because the metric becomes singular \cite{RosenPlane,EinsteinRosen}. 
 Later it was shown by I. Robinson and by H. Bondi  that there exists a coordinate transformation that removes Rosen's singularity, so in the geometrical theory, such waves are in fact
 nonsingular and acceptable \cite{BondiNature,BPR}. We suspect that the plane wave solution can be fit into the SRA by truncating it at the coordinate singularity, which would then occur at unrenormalized future infinity. A mathematical treatment of these matters should appear in the future.

  The SRA has the advantage of simple and fixed notion of causality at the classical level, because the flat null cone serves as a bound on the curved one.  In this view, one becomes less 
dependent on the study of  topology, global techniques, careful definitions of causality of various sorts, and the like, such as modern texts contain  \cite{HawkingEllis,Wald}.  The SRA is therefore simpler in an obvious way than the alternatives.  There is no difficulty in extending this nondynamical causal structure into
 the quantum regime, so there should be no problem
 in writing down equal-time commutation relations, \emph{etc.} in the usual way. 
 Thus,  worries expressed  by Ashtekar and Geroch, Isham, van Nieuwenhuizen, Wald and Weinstein above  are resolved.  

	While the special relativistic approach to  Einstein's equations is  \emph{locally} and \emph{classically} equivalent to the usual theory, as we saw earlier, there might be different \emph{global} or \emph{quantum} properties.  For example, though flat spacetime with trivial topology is stable in general relativity \cite{DeserChoquet}, closed flat space is unstable \cite{DeserBrill}, so it
 appears that the usual topology is more than just a simple and convenient choice for the SRA.  Also, it likely will turn out that some regions of spacetime in complete exact solutions of the geometrical theory simply 
do not exist in the SRA, perhaps due to infinite postponement from the lapse's tending toward $0.$  Moreover, the SRA of general relativity has less gauge freedom than 
the geometrical and field formulations, because any gauge choice that leads to an improper null cone relationship must be prohibited.  To be more precise, the SRA configurations form a proper subset of the naive configurations, but with the same order of infinite size.  The use of a new set of variables, in which only
 the null-cone respecting field configurations are possible, would be a way to prohibit them.  We propose such a set below.

	The attitude of regarding the flat spacetime as fundamental has been most visibly promoted by A. A. Logunov and colleagues \cite{LogunovFund,LogunovBasic,LogunovBook} (to name a few).\footnote{This school has also produced an energetic critique of geometrical general relativity as lacking physical meaning,
 in the sense of lacking conservation laws and failing to make definite predictions.  We do not endorse this critique.}
 To distinguish their view clearly from any geometrical notions, they have given the name ``relativistic theory of gravitation'' (RTG) to the work.  The nature of the RTG has evolved 
slightly over the years.  
For some time it consisted in Rosen's  tensorial $\Gamma\Gamma$ action for general relativity and his tensorial DeDonder condition \cite{Rosen1} postulated as necessary, presumably with specification of trivial topology for spacetime.  There is also attached a ``causality principle''
 that requires that the 
curved null cone not violate the flat one \cite{LogunovFund,Chugreev,Pinson}.  This causality principle, which seems to have appeared following criticisms by Zel'dovich and Grishchuk \cite{Zel2,Grishchuk90}, is
  the feature most relevant to our purposes.  More recently, the RTG has often featured acquired a mass term, but we are interested especially in the massless version.  (The massive version will have the same null cone consistency problem for gravitational waves.)

	Logunov \emph{et al.}, being committed to the flat spacetime view,
regard the question of compatible null cones as requiring a  solution.  
Furthermore, they believe it to be solved already by their causality principle, which we shall call the Logunov Causality Principle.  This principle states that field
configurations that
make the curved metric's null cone open wider than the flat metric's are
physically meaningless \cite{LogunovFund,Chugreev,Pinson}.  As they observe, satisfaction is
not guaranteed (even with their gauge conditions, notes Grishchuk \cite{Grishchuk90}), which means that the set of partial differential equations is not enough to define the theory.   The Logunov Causality Principle is therefore enforced ``by hand.''
Some causality principle is indeed needed, but the Logunov Causality Principle strikes
us as somewhat arbitrary and \emph{ad hoc}.  One would desire three improvements.  First, one would prefer that the causality principle be  tied somehow to the
Lagrangian density or field variables, not separately appended \cite{Grishchuk90}.  Second, one wants a guarantee that there exist enough solutions obeying the principle to cover all physically relevant situations.  Third, one would prefer a more
convenient set of variables to
describe the
physics.  We  address all of these matters below.

  Concerning the first shortcoming, it might be suggested \cite{LogunovBook}  that the Logunov causality principle is analogous to the energy conditions
 \cite{Wald} that one typically imposes. However, this analogy strikes us as weak.  The dissimilarity is in how the two conditions accept or reject solutions.  The energy conditions are used to exclude or 
include whole classes of matter fields, so any configuration with one sort of matter field--perhaps a minimally coupled massless scalar field with the correct sign in the Lagrangian density--
 is permitted, whereas any configuration with another sort of matter field--perhaps the scalar field with the wrong sign-- is prohibited as unphysical.   This criterion expresses the idea that
 some sorts of matter are physically reasonable, but others are not.  Furthermore, there is no worry that a permissible sort of matter could evolve into a forbidden sort in accord with the field equations.  On the other hand, the Logunov Causality Principle 
cannot give (or at least has not given) a similarly general explanation for why it rejects some solutions of the field equations. 

	A more serious problem is that it cannot give 
any assurance that it permits a sufficiently large number of solutions to cover all physical situations that arise.  \emph{A priori} there is no reason to believe that one can
 (partially) fix the gauge, and then still reject some solutions in the appropriate gauge as 
unphysical.  \emph{A posteriori} there seems to be good evidence that this worry
is serious.  Let us imagine a young man, Nicholas, playing  on a drum set.  Contemplating the motion of his arms and sticks, we may be confident that the traceless part of the second time derivative of his quadrupole moment, contracted with itself, is usually nonzero.  
But that means that Nicholas emits gravitational radiation, for this is just the formula for the average power radiated in general relativity, under suitable assumptions \cite{Wald}.
There will be anisotropic but roughly spherical waves of gravitational radiation diverging from Nicholas.  
Far away from him,  these waves will look approximately like plane waves obeying linearized gravity with the tensorial Hilbert gauge  condition.  The behavior of plane monochromatic single-polarization waves in linearized general relativity is well-known \cite{Ohanian}.  In this gauge, the two energy-carrying  polarizations both consist of alternately shrinking one transverse
 direction and stretching the other,
 while the time (lapse) and spatial propagation directions are unaffected \cite{Ohanian}. (Below we will analyze this linear solution using a generalized eigenvalue-eigenvector formalism.)  But the shrinking of one spatial eigenvalue while leaving the time lapse unaltered implies a 
violation of the flat 
metric's null cone.  A test particle passing through the wave could achieve superluminal speeds.  In short, it appears that, if the exact behavior of the plane waves is anything like the linearized behavior, then monochromatic gravitational radiation satisfying the tensorial
 DeDonder condition generically violates the Logunov causality principle.  While monochromatic radiation is a rather idealized case, and thus perhaps need not obey $\eta$-causality by itself for a satisfactory theory,
it is not at all obvious that the superposition 
of monochromatic waves which all individually violate $\eta$-causality yields a sufficiently generous set of realistic waves that satisfy that condition.  
Thus, there is reason to worry that the Logunov causality principle cannot be implemented, because Nicholas in fact can play drums.  Evidently, arranging for  wave solutions to obey the causality principle is rather more difficult than addressing most of the solutions that Logunov and collaborators have addressed to date, which often have considerable symmetry and fairly trivial dynamics.
\cite{LoskutovCosmos,Chugreev,Pinson,Ionescu}.  Thus, some way of enforcing null cone consistency without excluding necessary solutions of the field equations must be sought.  

	The lesson that we draw from the apparent shortcomings of the RTG in its present and past forms is not, \emph{pace} some authors  \cite{Zel1,Zel2,Grishchuk90}, that the flat metric must be considered merely a useful fiction.  Rather, 
if the SRA is to be maintained, then a  more fundamental approach to securing consistency between the null cones must be sought. 
 One will want to use the gauge freedom of general relativity to secure null cone consistency. In that way, one can be confident that a sufficiently large number of solutions exist, because one member from each equivalence class of solutions will be included.


\section{Describing and Enforcing the Proper Null Cone Relationship}

\subsection{Consistent Null Cones by Suitable Gauge Restrictions?}

     As several authors above pointed out, the local relation between the two null cones is  indeed gauge-dependent
in general relativity \cite{Penrose,Grishchuk90}.  (Below we will circumvent this problem by redefining gauge transformations.) One might therefore hope to design a set of
gauge-fixing conditions that yield the desired behavior, or at least to impose restrictions on the variables that exclude unsuitable gauge choices while permitting suitable ones. 

	To this end, the work of M. Visser, B. Bassett, and S. Liberati on ``superluminal censorship'' is quite interesting \cite{SuperCens,PertSuperCens}.  Working in linearized gravity in the usual DeDonder-Lorentz-Hilbert gauge, and excluding gravitational radiation, these authors find that the curved metric's null cone opens wider than the flat background's only if the null energy condition (NEC) is violated. The NEC being rather commonly satisfied, this result is quite encouraging for efforts to respect the flat metric's null cone in full nonlinear general relativity.  However, as they note, the NEC does not hold universally.  They also note that this result does not obviously or easily generalize to strong field situations.  Finally, we note that their exclusion of gravitational radiation is a severe limitation, for, as the study of gravitational waves in linearized general relativity shows, gravitational radiation in this gauge \emph{does} result in widening of the null cone relative to that of the flat background.  That is why we will renounce this gauge condition, so either a better gauge must be found, or (more likely) one must give up gauge-fixing.  Rather than putting
any gauge conditions into the theory  by hand, one would prefer to implement them in the action principle or the choice of field variables somehow.

	It might be hoped that the ADM split of the metric \cite{MTW,Wald}, which is quite useful in applications and in identifying the true degrees of freedom, 
would be a good language for discussing the null cone consistency issue.  Let us see if that is the case.  For convenience we choose Cartesian coordinates, so that $\eta_{\mu\nu} = diag(-1,1,1,1).$  We therefore make an ADM split of Logunov's 4-dimensional analysis of the causality principle.
In considering whether all the
 vectors $V^{\mu}$ lying on $\eta$'s null cone are $g$-timelike, $g$-null, or $g$-spacelike, it suffices to consider future-pointing vectors with unit time component; thus $V^{\mu} = (1, V^{i})$, where $V^{i} V^{i} = 1 $ (the sum running
from 1 to 3).  The causality principle can be written $h_{ij} (\beta^{i} + V^{i}) (\beta^{j} + V^{j}) - N^{2}   
\geq  0$ for all spatial unit vectors $V^{i}$.  Here the spatial metric is $h_{ij}$, 
the lapse is $N$, and the shift is $\beta^{i}$.  One could visualize this
equation as the requirement that an 
ellipsoid (not centered at the origin if the shift $\beta^{i}$ is nonzero) not 
protrude from the closed unit ball.
Unfortunately, the  ``all''  in ``for all spatial unit vectors'' is not too easy to handle, so we will in fact look for a better language than an ADM split for discussing null cone consistency.
	
	If there to be any is hope for restricting the gauge freedom so as to ensure that the curved null cone stays consistent with the flat one, then there must be ``enough'' gauge freedom to transform any
 physically significant solution into a form that satisfies $\eta$-causality. It is (very roughly) the case that the curved spatial metric controls the
width of the light cone, while the shift vector determines its tilt from the vertical (future) direction and the
lapse function determines its length.  For general relativity, the spatial metric contains the physical degrees of freedom; the lapse and shift represent
the gauge freedom, so they can be chosen largely arbitrarily.
By analogy with conditions typically imposed in geometrical general
relativity to avoid causality difficulties \cite{Wald}, one would prefer, if possible,
that the curved light cone be \emph{strictly inside} the flat light cone, not tangent to it, because tangency indicates that the field is on the verge of $\eta$-causality violation. This requirement we call ``stable $\eta$-causality,'' by analogy to the usual condition of stable causality \cite{Wald}.  One might worry that this requirement would exclude all curved metrics conformally related to the flat one,
 and even the presumed ``vacuum'' $g_{\mu\nu}=\eta_{\mu\nu}$ itself.  Indeed it does, but 
if one takes the message of gauge invariance seriously, then there is no fundamental basis for preferring $g_{\mu\nu}=\eta_{\mu\nu}$ over having the curved metric agree 
\emph{up to a gauge transformation} with the flat metric. 

	Let the desired relation between the null cones hold at some initial moment.  Also let the curved spatial metric and shift be such at some event in that moment that they tend to make the curved cone violate the flat one a bit later.
By suitably reducing the lapse, one can lengthen the curved cone until it once again is safely inside the flat cone.  By so choosing the lapse at all times and places, one should be able to
satisfy the causality principle at every event, if no global difficulties arise.
In a rough sense, one might use up $\frac{1}{4}$ of the gauge freedom of general relativity, while leaving the remainder.  It is possible that this procedure forces the lapse toward $0$ in some cases, which implies that physical events are stretched out over more and more of Minkowski spacetime, perhaps to future or past infinity.  (This procedure bears a formal resemblance to the use of singularity-avoiding coordinates in
 numerical relativity \cite{Lehner}, in which the gauge (coordinate) freedom is used to exile singularities to infinite coordinate values.  The obvious difference is that here the trick is suitably invariant.)  

	Frequently it is assumed that the reason that the gravitational Hilbert action is gauge-invariant is because such gauge invariance reflects a deep feature of the world, general covariance.  However, as we saw above, one can give a somewhat humbler explanation: it is known from the flat spacetime approach that eliminating the time-space components of the field is essential for positive energy 

properties in Lorentz-invariant theories \cite{Fierz,van Nieuwenhuizen}, though in fact the time-time component need not be \cite{Cavalleri}.  
We might suggest that the gauging away of the time-time component is necessary rather to respect $\eta$-causality.  So gauge invariance appears to be required to respect positive energy and special relativistic causality.

\subsection{The Causality Principle and Loose Inequalities}

	As should be clear from the worries about conformally flat curved metrics, the desired relationship between the two null cones takes the form of some loose inequalities $a \leq b$.  Such relations have been called ``unilateral''
 \cite{OdenReddy,Del Piero 1,Del Piero 2,MoreauPS,Marques} or ``one-sided'' \cite{Lacomba,Ibort,Cantrijn},  typical examples being nonpenetration conditions.  Such constraints are rather more difficult to
 handle than the standard ``bilateral'' or ``two-sided'' constraint \emph{equations} that most treatments of constraints in physics discuss. 
Loose inequalities are also more difficult to handle than strict inequalities $a < b,$ such as the  positivity conditions in canonical general relativity \cite{KlotzDiss,GoldbergKlotz,IshamKakas,Kuchar92,Klauder2}, which require that the ``spatial'' metric be spatial.  One might
 eliminate the positivity conditions by a change of variables \cite{GoldbergKlotz} that satisfies the inequalities identically, such as an exponential function  $h = e^{y},$ as Klotz contemplates.   

	If one does leave causality constraints in the theory, rather than solving them as was suggested in the previous paragraph, then one must worry about impulsive constraint forces, unless one makes the constraint ``ineffective'', so the constraint force vanishes on account of the constraint itself \cite{PSShepley}.  However, it seems considerably more satisfactory to reduce the configuration space of the theory so that the problem is avoided altogether. 

\subsection{New Variables and the Segr\'{e} Classification of the Curved Metric with Respect to the Flat}

	Previous formulations of the causality principle, which have used the
metric \cite{LogunovBasic,LogunovFund,Chugreev,Pinson} or ADM variables as above, have been sufficiently inconvenient to render
progress difficult.  This was our third complaint about Logunov's formulation of $\eta$-causality.  
 One could achieve a slight savings by using the conformally invariant weight $-\frac{1}{2}$ densitized part of the metric $ g_{\mu\nu} (-g)^{-\frac{1}{4}}.$  Then 
nine numbers at each event are 
required (the determinant being $-1$), which is
a bit better than the 10 of the full metric, but still too many. 

  One would like to diagonalize
$g_{\mu\nu}$ and $\eta_{\mu\nu}$ simultaneously by solving the generalized eigenvalue problem 
\begin{equation}
 g_{\mu\nu} V^{\mu} = \Lambda \eta_{\mu\nu} V^{\mu},
\end{equation}
or perhaps the related problem using $ g_{\mu\nu} (-g)^{-\frac{1}{4}}.$  
   However, in general that is
impossible, because there is not a complete set
of eigenvectors, due to the indefinite (Lorentzian) nature of both tensors \cite{Eisenhart,HallArab,HallDiff,HallNegm,Hall5d,BonaCollMorales} (and references in (\cite{BonaCollMorales}).
There are four 
Segr\'{e} types for a real symmetric rank 2 tensor with respect to a Lorentzian metric,
the several types having 
different numbers and sorts of
eigenvectors \cite{HallArab,HallDiff,HallNegm,Hall5d}. 

 To our knowledge, the only previous work to consider a generalized eigenvector decomposition of a curved Lorentzian metric with respect to a flat one\footnote{There is also a literature on choosing coordinates to diagonalize a curved metric \cite{Tod}.  According to K. P. Tod, ``there is very little change for Lorentzian metrics'' compared to Riemannian
 ones \cite{Tod}.  But there is a large change for the eigenvalue problem of interest to us in changing from a Riemannian to a Lorentzian background metric, so the connection between these problems must be somewhat loose.} was done by  I.
Goldman \cite{Goldman}, in the context of Rosen's bimetric theory
of gravity, which does not use Einstein's field equations.
Goldman's work was tied
essentially to Rosen's theory, so it does not address our concerns much.   The lack of gauge freedom in Rosen's theory also ensured that the curved null cone was not
 subject to adjustment, unlike the situation in general relativity with a flat metric.  As it happens, in Rosen's theory, 
for static spherically
symmetric geometries, the causality principle is always \emph{violated}\footnote{Prof. Goldman has kindly provided this information from his dissertation in Hebrew.}, so we conclude that  the theory is not consistent with special relativity.  However, Einstein's theory evidently has enough gauge freedom to make a special relativistic approach possible.

	An eigenvalue-eigenvector decomposition for the \emph{spatial} metric was briefly contemplated by Klotz and Goldberg \cite{KlotzDiss,GoldbergKlotz}.  For space, 
as opposed to spacetime, one has a positive definite background (identity) matrix, so the usual theorems apply. But Klotz and Goldberg, who did not assume a flat metric tensor to exist,
 found little use for the eigenvector decomposition because of the nontensorial nature of the $3 \times 3$ identity matrix.  Such a decomposition, even given a flat metric tensor, is still somewhat complicated if the ADM shift is nonvanishing ($g_{0i} \neq 0$), as it usually is. 
 Diagonalization has been quite useful in the study of spatially homogeneous cosmologies \cite{Jantzen}, but our interest is not in specific solutions only, but the general case.
	
	Let us now proceed with the diagonalization project, confining our attention to four spacetime dimensions.  (A brief assertion without proof of a few results from the eigenvector formalism appeared  recently \cite{NCTRA}.)  Given that a complete set of generalized eigenvectors might fail to exist, it is necessary to consider how many eigenvectors do exist and
under which conditions. This problem has been substantially addressed in a different context by G.
S. Hall and collaborators \cite{HallArab,HallDiff,HallNegm,Hall5d}, who were
interested in classifying the stress-energy  or Ricci tensors with respect to
the (curved) metric in (geometrical) general relativity. Such problems have in fact been studied over quite a long period of time \cite{BonaCollMorales} (and references therein), but we find the work of Hall \emph{et al.} to be especially convenient for our purposes.   
 There exist four cases, corresponding to the four
possible Segr\'{e} types (apart from degeneracies)  for the classified tensor.
The case $\{1,111\}$ has a complete set of eigenvectors ($1$ timelike, $3$
spacelike with respect to $\eta$), and is thus the most convenient case.  The
case  $\{211\}$ has two spacelike eigenvectors and one null eigenvector (with
respect to $\eta$), whereas the  $\{31\}$ case has 
one spacelike eigenvector and one null one.  The last case, $\{z\,\bar{z}11\}$   has 2  spacelike eigenvectors with real eigenvalues and 2 complex eigenvalues.

	We now consider the conditions under which metrics of each of these Segr\'{e} classes obey $\eta$-causality.  To give a preview of our results, we state
that the $\{1,111\}$ and $\{211\}$ cases sometimes do obey it, although the
$\{211\}$  metrics appear to be dispensable.  But no metric of type $\{31\}$ or
$\{z\,\bar{z}11\}$  obeys
 the causality principle, so these types can be
excluded from consideration for the SRA.  

 	Hall \emph{et al.} introduce a real null tetrad of vectors $L^{\mu}, N^{\mu}, X^{\mu}, Y^{\mu}$ 
with vanishing inner products, apart from the relations $\eta_{\mu\nu} L^{\mu}N^{\nu} =
\eta_{\mu\nu} X^{\mu} X^{\nu} = \eta_{\mu\nu} Y^{\mu} Y^{\nu} = 1$, so $L^{\mu}$ and $N^{\mu}$ are null, while
$X^{\mu}$ and $Y^{\mu}$ are spacelike. 
 (The signature is $-+++$.)  Expanding
an arbitrary vector $V^{\mu}$ as $V^{\mu} = V^{L} L^{\mu} + V^{N} N^{\mu} +
V^{X} X^{\mu} + V^{Y} Y^{\mu}$ and taking the $\eta$-inner product with each vector
of the null tetrad reveals that $V^{L} = \eta_{\mu\nu} V^{\mu} N^{\nu} $, $V^{N} = \eta_{\mu\nu} V^{\mu} L^{\nu}$,
$V^{X} = \eta_{\mu\nu} V^{\mu} X^{\nu} $, and $V^{Y} = \eta_{\mu\nu} V^{\mu} Y^{\nu}$.  
Thus, the Kronecker delta
tensor can be written as $\delta^{\mu}_{\nu} = L^{\mu} N_{\nu} + L_{\nu}
N^{\mu} + X^{\mu} X_{\nu} + Y^{\mu} Y_{\nu},$ indices being lowered here  using $\eta_{\mu\nu}$.  For some purposes it is also convenient to
define the timelike vector $T^{\mu} = \frac{L^{\mu} - N^{\mu} }{\sqrt{2} }$ and  the spacelike vector $Z^{\mu} = \frac{ L^{\mu} + N^{\mu} }{\sqrt{2}}$.

	We employ the results of Hall \emph{et al.} \cite{HallArab,HallDiff,HallNegm,Hall5d}, who find that 
the four possible  Segr\'{e} types (ignoring degeneracies) for a
(real) symmetric rank 2 tensor in a four-dimensional spacetime with a Lorentzian
metric can be written in the following ways, using a well-chosen
null tetrad.  The type $\{1,111\}$ can be written as
\begin{eqnarray}
g_{\mu\nu} = 2\rho_{0} L_{(\mu} N_{\nu)} + \rho_1 (L_{\mu} L_{\nu} + N_{\mu} N_{\nu} )
+ \rho_2 X_{\mu} X_{\nu} + \rho_3 Y_{\mu} Y_{\nu},
\end{eqnarray} or equivalently
\begin{eqnarray}
g_{\mu\nu} = -(\rho_{0} - \rho_{1} ) T_{\mu} T_{\nu} +  (\rho_{0} + \rho_{1} ) Z_{\mu} Z_{\nu} + 
\rho_2 X_{\mu} X_{\nu} + \rho_3 Y_{\mu} Y_{\nu}.
\end{eqnarray}
As usual, the parentheses around indices mean that the symmetric part should be taken \cite{Wald}.
The type $\{211\}$ can be written as 
\begin{eqnarray}
g_{\mu\nu} = 2\rho_{1} L_{(\mu} N_{\nu)} + \lambda L_{\mu} L_{\nu} 
+ \rho_2 X_{\mu} X_{\nu} + \rho_3 Y_{\mu} Y_{\nu}, 
\end{eqnarray} with $\lambda \neq 0$, the null eigenvector being $L^{\mu}$.
The type $\{31\}$ can be written as 
\begin{eqnarray}
g_{\mu\nu} = 2\rho_{1} L_{(\mu} N_{\nu)}  + 2 L_{(\mu} X_{\nu)}  
+ \rho_1 X_{\mu} X_{\nu} + \rho_2 Y_{\mu} Y_{\nu},  
\end{eqnarray} 
the null eigenvector again being $L^{\mu}$.
 The final type, $\{z\,\bar{z}11\}$, can be written as
\begin{eqnarray}
g_{\mu\nu} = 2\rho_{0} L_{(\mu} N_{\nu)} + \rho_1 (L_{\mu} L_{\nu} - N_{\mu} N_{\nu} )
+ \rho_2 X_{\mu} X_{\nu} + \rho_3 Y_{\mu} Y_{\nu},
\end{eqnarray} with $\rho_1 \neq 0$.
The requirements to be imposed upon the curved metric  for the moment are the following:  all $\eta$-null vectors must be $g$-null or $g$-spacelike, all $\eta$-spacelike eigenvectors must be $g$-spacelike, $g_{\mu\nu}$ must be 
Lorentzian (which amounts to having a negative determinant), and $g_{\mu\nu}$ must be connected to $\eta_{\mu\nu}$ by a succession of small changes which respect $\eta$-causality and the Lorentzian signature.  
It convenient to employ a slightly redundant form that admits all four types in order to treat them simultaneously.  Thus, we write
\begin{eqnarray}
g_{\mu\nu} = 2 A L_{(\mu} N_{\nu)} + B L_{\mu} L_{\nu} +  C N_{\mu} N_{\nu} 
+ D X_{\mu} X_{\nu} + E Y_{\mu} Y_{\nu} + 2 F  L_{(\mu} X_{\nu)}.
\end{eqnarray}  Using this form for $g_{\mu\nu}$, one
readily finds the squared length of a vector $V^{\mu}$ to be
\begin{eqnarray}
g_{\mu\nu} V^{\mu}V^{\nu} = 2 A V^{L} V^{N} + B (V^{N})^{2} + C (V^{L})^{2} + D (V^{X})^{2} 
+ E (V^{Y})^{2} \nonumber \\  + 2 F V^{X} V^{N}.
\end{eqnarray}
It is not clear \emph{a priori} how to express sufficient conditions for the causality principle 
in a convenient way.  
But it will turn out that the necessary conditions that we can readily impose are also sufficient.

\subsection{Necessary Conditions for Respecting the Flat Metric's Null Cone}

	The causality principle requires that the $\eta$-null vectors $L^{\mu}$ and $N^{\mu}$ be $g$-null or $g$-spacelike, 
so $B \geq 0, C \geq 0$.  These conditions already exclude the type   
$\{z\,\bar{z}11\}$, because the form above 
 requires that $B$ and $C$ differ in sign.  It must also be the case that the $\eta$-spacelike vectors 
$X^{\mu}$ and $Y^{\mu} $ are $g$-spacelike, so  $D>0$ and $E>0$.  

	Not merely  $L^{\mu}$ and $N^{\mu}$, but all $\eta$-null vectors must be $g$-null or $g$-spacelike. 
 This requirement quickly implies that $E \geq A$, and also requires that 
\begin{eqnarray}
 B (V^{N})^{2} + 2 F V^{X} V^{N} + (D-A) (V^{X})^{2} \geq 0.
\end{eqnarray} 
Here there are two cases to consider:  $F=1$ for type $\{31\},$ and $F=0$ for types  $\{1,111\}$ and $\{211\}.$  Let us consider $F=1.$  The $\{31\}$ has $B=0$, so the equation reduces to $2 F V^{X} V^{N} + (D-A) (V^{X})^{2} \geq 0, $ which implies that either $ V^{X} = 0$ or, failing that, $2 F V^{N} + (D-A) V^{X} \geq 0. $   Clearly one could also consider a null vector with the opposite value of $V^{X},$ 
yielding the inequality   $2 F V^{N} - (D-A) V^{X} \geq 0. $  Adding these two inequalities gives $4 V^{N} \geq 0,$ which simply cannot be made to hold for all values of $V^{N}.$  Thus, the $F=1$ case yields no $\eta$-causality-obeying curved metrics, and the   $\{31\}$ type is eliminated.  It remains to consider $F=0$ for the $\{1,111\}$ and $\{211\}$ types.  The resulting inequality is 
$B (V^{N})^{2}  + (D-A) (V^{X})^{2} \geq 0.$  Because $B \geq 0$ has already been imposed, it follows only that $D \geq A.$

	Let us summarize the results so far.  The inequalities  $B \geq 0$ and $C \geq 0$ have excluded the type $\{z\,\bar{z}11\}.$  We also have $D > 0,$ $D \geq A,$ $E > 0,$ $ E \geq A.$ Finally, $F=0$ excludes the type $\{31\},$ so only $\{1,111\}$ and $\{211\}$ remain. 

	We now impose the requirement of Lorentzian signature.  At a given event, one can find a  coordinate $x$ such that $ (\frac{\partial}{\partial x})^{\mu} = X^{\mu}$ and (flipping the sign of $Y^{\mu}$ if needed for the orientation)  a coordinate $y$ such that $ (\frac{\partial}{\partial y})^{\mu} = Y^{\mu};$ these two coordinates can be regarded as Cartesian.  Then the null vectors $L^{\mu}$ and $N^{\mu}$ lie in the $t-z$ plane 
of this sort of Cartesian system.  The curved metric has a block diagonal part in the $x-y$ plane with positive determinant, so imposing a Lorentzian signature means ensuring a negative determinant for the $2 \times 2$ $t-z$ part.  The vectors $L^{\mu}$ and $N^{\mu}$ in one of these coordinate systems take the form $L^{\mu} = (L^{0}, 0,0,L^{3})$ and $N^{\mu} = (N^{0}, 0 , 0, N^{3}).$  Given
the Cartesian form $\eta_{\mu\nu} = diag(-1,1,1,1)$ and the nullity of these two vectors, it follows that $|L^{0}| = |L^{3}|$ and $|N^{0}| = |N^{3}|.$  Therefore the relevant parts of the curved metric can be written in such a coordinate basis as 
\begin{eqnarray}
g_{00} = 2 A L^{0} N^{0} + B (L^{0})^{2} + C (N^{0})^{2},  \nonumber \\
g_{03}=g_{30}= -A(N^{0} L^{3} + L^{0} N^{3}) - B L^{0} L^{3} - C N^{0} N^{3}, \nonumber \\
g_{33} = 	2 A L^{3} N^{3} + B (L^{3})^{2} + C (N^{3})^{2}. 
\end{eqnarray}  
Taking the determinant using \emph{Mathematica} and recalling that  $|L^{0}| = |L^{3}|$ and $|N^{0}| = |N^{3}|,$ one finds that the condition for a negative determinant is $2(A^{2} - BC) |L^{3}|^{2} |N^{3}|^{2} (sign(L^{0} L^{3} N^{0} N^{3}) -1) < 0.$  The linear independence of $L^{\mu}$ and $N^{\mu}$ implies that $sign(L^{0} L^{3} N^{0} N^{3}) = -1,$ so the determinant condition is $A^{2} -BC >0.$  Because $B$ and $C$ are both nonnegative,  $A^{2} -BC >0$  implies that $A \neq  0.$  But the requirement that the curved metric be smoothly deformable through a sequence of signature-preserving steps  means that the curved metric's value of $A$ cannot ``jump'' from one sign of $A$ to another, but must agree with the flat metric's positive sign.  It follows that $A > 0.$  

	We now summarize the necessary conditions imposed:
\begin{eqnarray}
A>0,  & A^{2}>BC, & B\geq 0, \nonumber \\ C \geq 0, & D \geq A, & E \geq A, \nonumber \\ & F=0. 
\end{eqnarray}

\subsection{Sufficient Conditions for Respecting the Flat Metric's Null Cone}

	Thus far, it is not clear whether these necessary conditions are sufficient.  We now prove that they are.  It is helpful to consider the two types, $\{1,111\}$ and $\{211\}$, separately.

	For the type $\{1,111\}$, the conditions on the coefficients $A,B,$ \emph{etc.} reduce to 
\begin{eqnarray}
A>0, & A>B, & B\geq 0, \nonumber \\ C = B, & D \geq A, & E \geq A.
\end{eqnarray}  For this form the following relations between variables hold:
\begin{eqnarray} A = \rho_{0}, & B = \rho_{1}, & D = \rho_{2},  \nonumber \\ & E = \rho_{3}.
\end{eqnarray}  It follows that this type can be expressed as 
\begin{eqnarray}
g_{\mu\nu} = -(A-B) T_{\mu} T_{\nu} +  (A + B) Z_{\mu} Z_{\nu} + 
D X_{\mu} X_{\nu} + E Y_{\mu} Y_{\nu}. 
\end{eqnarray}
Writing the eigenvalues for $T^{\mu},$ $X^{\mu},$ $Y^{\mu},$ and $Z^{\mu}$ as $D^{0}_{0},$ $D^{1}_{1},$ $D^{2}_{2},$ and $D^{3}_{3}, $ respectively, one has 
\begin{eqnarray}
D^{0}_{0} = A-B, & D^{1}_{1} = A + B,  & D^{2}_{2} = D, \nonumber \\
& D^{3}_{3} = E.
\end{eqnarray}
One sees that the inequalities imply that the eigenvalue for the timelike eigenvector $T^{\mu}$  (briefly, the ``timelike eigenvalue'') is no larger than any of the spacelike eigenvalues:
\begin{eqnarray}
D^{0}_{0} \leq D^{1}_{1}, & D^{0}_{0} \leq D^{2}_{2}, &  D^{0}_{0} \leq D^{3}_{3}, 
\end{eqnarray}
and that all the (generalized) eigenvalues are positive.
Let us now see that these conditions are sufficient.  Writing an arbitrary vector $V^{\mu}$ as $V^{\mu} = V^{T} T^{\mu} + V^{X} X^{\mu} + V^{Y} Y^{\mu} + V^{Z} Z^{\mu},$ one sees that its $\eta$-length (squared) is $\eta_{\mu\nu} V^{\mu}V^{\nu} = -(V^{T})^{2} + (V^{X})^{2} + (V^{Y})^{2} + (V^{Z})^{2} .$  Clearly this length
 is never more positive than $\frac{1}{D^{0}_{0} }  g_{\mu\nu} V^{\mu}V^{\nu} = -(V^{T})^{2} + \frac{ D^{1}_{1} }{D^{0}_{0} } (V^{X})^{2} + \frac{ D^{2}_{2} }{D^{0}_{0} }  (V^{Y})^{2} + \frac{ D^{3}_{3} }{D^{0}_{0} }  (V^{Z})^{2}, $ so the necessary conditions are indeed sufficient for type    $\{1,111\}.$

	For the type $\{211\}$, the conditions on the coefficients $A,B,$ \emph{etc.} reduce to 
\begin{eqnarray}
A>0, & B>0, & C = 0, \nonumber \\ D \geq A, & E \geq A, & F=0.
\end{eqnarray}
One can write the curved metric in terms of $T^{\mu}$, $Z^{\mu}$, $X^{\mu}$, and $Y^{\mu}$, 
though $T^{\mu}$ and $Z^{\mu}$ are not eigenvectors.  One then has
\begin{eqnarray}
g_{\mu\nu} = -(A - \frac{1}{2}B) T_{\mu} T_{\nu} +  (A + \frac{1}{2}B ) Z_{\mu} Z_{\nu}  + B Z_{(\mu} T_{\nu)}    +  D X_{\mu} X_{\nu} 
+ E Y_{\mu} Y_{\nu}. \end{eqnarray}
Writing an arbitrary $\eta$-spacelike vector field $V^{\mu}$ as $V^{\mu} = G T^{\mu} + H Z^{\mu} + I X^{\mu} + J Y^{\mu}$,
 with $H^{2} + I^{2} + J^{2} > G^{2}$, one readily finds the form of
$g_{\mu\nu} V^{\mu} V^{\nu} $.  Employing the relevant inequalities and shuffling coefficients, one
obtains the manifestly positive result $g_{\mu\nu} V^{\mu} V^{\nu} = A(H^{2} + I^{2} + J^{2} -
G^{2}) + \frac{1}{2}B(G-H)^{2} + (D-A)I^{2} + (E-A) J^{2}$.  This positivity
result says that all $\eta$-spacelike vectors are $g$-spacelike.  Earlier the
requirement that all $\eta$-null vectors be $g$-null or $g$-spacelike was
imposed.  These two conditions together comprise the causality principle, so
we have obtained sufficient conditions for the $\{211\}$ type also.  

	The $\{211\}$ type, which has with one null and two spacelike eigenvectors,  is a borderline case in which the curved metric's null cone is tangent to the flat metric's cone along a single direction \cite{ChurchillVector}. 
 Clearly such borderline cases of $\{211\}$ metrics obeying the causality principle form in some sense  a measure $0$ set of all causality principle-satisfying metrics. 
 Given that they are so scarce, one might consider neglecting them.  Furthermore, they are arbitrarily close to violating the causality principle.  We recall the 
 criterion of stable causality in geometrical general relativity \cite{Wald}  (where the issue is closed timelike curves, without regard to any flat metric's null cone),
 which frowns upon metrics which satisfy causality, but would fail to do so if
 perturbed by an arbitrarily small amount.  One could imagine  that quantum fluctuations might push such a marginal metric over the edge, and thus prefers to exclude such metrics as unphysical.
  By analogy, one might impose stable $\eta$-causality, which excludes curved metrics that are arbitrarily close to violating the flat null cone's 
notion of causality, though we saw that such a condition would exclude conformally flat metrics, also.  
Perhaps a better reason for neglecting type $\{211\}$ metrics is that they are both  technically inconvenient and physically unnecessary.  Because $\eta$-causality-respecting $\{211\}$ metrics are arbitrarily close to $\{1,111\}$ metrics, one could merely make a small gauge 
transformation
to shrink the lapse a bit more and obtain a $\{1,111\}$ metric instead.    Thus, every curved metric that respects $\eta$-causality either is of type $\{1,111\},$ or is arbitrarily close to being of type  $\{1,111\}$  and deformable thereto by a small gauge transformation reducing the lapse.  

	It follows that there is no loss of generality in restricting the configuration space to type  $\{1,111\}$  curved metrics,   for which the two metrics are simultaneously diagonalizable.  As a result, there exists a close relationship between 
the traditional orthonormal tetrad formalism and this eigenvector decomposition. 
 In particular, one can build a $g$-orthonormal tetrad field $e^{\mu}_{A}$ simply by choosing the normalization of the eigenvectors. This choice fixes  the local Lorentz freedom of the tetrad (except
 when eigenvalues are degenerate) in terms of the flat metric tensor. 

 Rewriting the generalized eigenvector equation for the case in which a complete set exists, one
 can write $g_{\mu\nu} e^{\mu}_{A} = \eta_{\mu\nu} e^{\mu}_{B} D^{B}_{A},$ with the four eigenvalues being the elements of the diagonal matrix $D^{A}_{B}$.  It is sometimes convenient to raise or lower the indices of this matrix using the matrix $\eta_{AB} = diag(-1,1,1,1)$.  
The tetrad field has $\{  e^{\mu}_{A} \}$ has  
inverse $\{  f_{\mu}^{A} \}$.    We recall the standard relations $g_{\mu\nu} = f_{\mu}^{A} \eta_{AB} f_{\nu}^{B} $ and   $g_{\mu\nu} e^{\mu}_{A} e^{\nu}_{B} = \eta_{AB}.  $
 It is not difficult to show the how the tetrad lengths are related to the eigenvalues:  $\eta_{\mu\nu} e^{\mu}_{A} e^{\nu}_{B} = D^{-1}_{AB},$ and equivalently, $\eta^{\mu\nu} f_{\mu}^{A} f_{\nu}^{B} = D^{AB}.$  It follows that 
 $f_{\nu}^{A}  = \eta_{\nu\alpha}  e^{\alpha}_{B} D^{AB}$, which says that a given leg of the cotetrad $f_{\nu}^{A}$ can be expressed solely in terms of the corresponding leg of the tetrad $e^{\mu}_{A}$, through a stretching, an index lowering, and possibly a sign change, without reference to the other legs.
Simultaneous diagonalization  implies that the tetrad vectors are orthogonal to each other with respect to \emph{both} metrics.


 \subsection{Linearized Plane Waves a Difficulty for the Logunov Causality Principle}

	As was stated in connection with Nicholas's drumming, monochromatic plane gravitational waves satisfying the  linearization of the Einstein equations and the Hilbert (linearized  DeDonder) 
gauge appear to violate   $\eta$-causality in general.  We will now show that in more detail, using Ohanian and Ruffini \cite{Ohanian} as our guide, while making use of the eigenvalue technology introduced above.  These results will also hold approximately for the Maheshwari-Logunov massive theory for large frequencies and weak fields.

	Defining a trace-reversed potential $\phi^{\mu\nu} = \gamma^{\mu\nu}
-\frac{1}{2} \eta^{\mu\nu}  \gamma$ (with $\gamma_{\mu}^{\mu} = \gamma$) and imposing the Hilbert gauge $\partial_{\mu} \phi^{\mu\nu} = 0,$ one puts the linearized Einstein equations in the
 form $\partial^{2} \phi^{\mu\nu} = 0.$  Because we desire plane wave solutions, let  $\lambda \phi^{\mu\nu} = H \epsilon^{\mu\nu} \cos(k_{\alpha} x^{\alpha} + \psi)$, where $\epsilon^{\mu\nu}$ is a constant polarization tensor,  $k^{\alpha}$ a constant polarization vector,
 and $H$ is a small number fixing the amplitude. We let the waves travel in the $z$-direction, so $k^{\mu} = \omega(1,0,0,1).$  We also define the vectors $\epsilon^{\mu}_{1} = (0,1,0,0)$ and $\epsilon^{\mu}_{2} = (0,0,1,0).$

	The gauge condition implies that the polarization tensor is orthogonal to the propagation vector: $ \epsilon^{\mu\nu} k_{\mu} = 0,$ leaving six independent solutions.  One can take the six independent polarization tensors (with both indices lowered using the flat metric) to be:
\begin{eqnarray}
\epsilon_{1\mu\nu} = \epsilon_{1\mu} \epsilon_{1\nu} - \epsilon_{2\mu} \epsilon_{2\nu}
=  \left[ \begin{array}{rrrr} 0 & 0 & 0 & 0 \\ 0 & 1 & 0 & 0 \\ 0 & 0 & -1 & 0 \\
0 & 0 & 0 & 0 \end{array} \right], \\
\epsilon_{2\mu\nu}=  \epsilon_{1\mu} \epsilon_{2\nu} + \epsilon_{2\mu} \epsilon_{1\nu}
=  \left[ \begin{array}{rrrr} 0 & 0 & 0 & 0 \\ 0 & 0 & 1 & 0 \\ 0 & 1 & 0 & 0 \\
0 & 0 & 0 & 0 \end{array} \right], \\
\epsilon_{3\mu\nu}=   \epsilon_{1\mu} \frac{1}{\omega} k_{\nu} + \epsilon_{1\nu} \frac{1}{\omega} k_{\mu}
=  \left[ \begin{array}{rrrr} 0 & -1 & 0 & 0 \\ -1 & 0 & 0 & 1  \\ 0 & 0 & 0 & 0 \\
0 & 1 & 0 & 0 \end{array} \right], \\
\epsilon_{4\mu\nu}=   \epsilon_{2\mu} \frac{1}{\omega} k_{\nu} + \epsilon_{2\nu} \frac{1}{\omega} k_{\mu}
=  \left[ \begin{array}{rrrr} 0 & 0 & -1 & 0 \\ 0 & 0 & 0 & 0  \\ -1 & 0 & 0 & 1 \\
0 & 0 & 1  & 0 \end{array} \right], \\
\epsilon_{5\mu\nu}=   k_{\mu} k_{\nu} =  \left[ \begin{array}{rrrr} 1 & 0 & 0 & -1  \\ 0 & 0 & 0 & 0  \\ 0 & 0 & 0 & 0 \\
-1 & 0 & 0  & 1 \end{array} \right],  \\
\epsilon_{6\mu\nu}=    \epsilon_{1\mu} \epsilon_{1\nu} + \epsilon_{2\mu} \epsilon_{2\nu}
=  \left[ \begin{array}{rrrr} 0 & 0 & 0 & 0  \\ 0 & 1 & 0 & 0  \\ 0 & 0 &  1 & 0 \\
0 & 0 & 0  & 0 \end{array} \right].
\end{eqnarray}

	The first two, which are transverse-transverse and traceless, are the physical (energy-carrying) polarizations in the massless theory.  One can show that they induce effective curved metrics of type $\{1111\},$ but,  partly due 
 to the oscillations 
of the cosine function, they violate $\eta$-causality, as the behavior of the eigenvalues shows.  The vector $U^{\mu} = (1,1,0,0)$ is $\eta$-null, but is $g$-timelike during every other half-period of the cosine function. 
Thus, this physical polarization violates $\eta$-causality.  The gauge waves do not seem helpful, because they all violate $\eta$-causality, too.

 \subsection{Dynamics of the Causality Principle}

	Above it was shown that the tensorial DeDonder gauge, which for many purposes is the best gauge choice possible, leads to causality violations.  It is doubtful that any better gauge choice exists, as far as the null cones are concerned, unless one gives up the condition that $
g_{\mu\nu} \rightarrow \eta_{\mu\nu} $ for weak fields.  But this condition is not gauge-invariant, so giving it up is not so difficult after all.  To achieve null cone consistency, we renounce gauge-fixing, \emph{pace} many authors (including ourselves!) \cite{Rosen1,Papapetrou,VlasovGaugeFix,Nikolic,SliBimGRG,NCTRA}. 
  
	Imposing stable $\eta$-causality gives the \emph{strict} inequalities
\begin{equation}
\bar{D}^{1}_{1} > 1, \bar{D}^{2}_{2} > 1, \bar{D}^{3}_{3} > 1. 
\end{equation} 
But strict inequalities can be solved, as Klotz suggested \cite{KlotzDiss,GoldbergKlotz}, and then the causality constraints would be eliminated.
This goal is achieved by  setting, for example,  
\begin{equation}
\bar{D}_{1}^{1} = e^{\alpha} + 1, \bar{D}_{2}^{2} = e^{\beta} + 1, \bar{D}_{3}^{3} = e^{\gamma} + 1.  
\end{equation}
 One could use these new variables to satisfy the causality constraints automatically.

 \subsection{Finite Gauge Transformations and Orthonormal Tetrads}

	The form of a \emph{finite} gauge transformation for the densitized inverse metric tensor in the field formulation is known from the work of Grishchuk, Petrov, and A. D. Popova \cite{Grishchuk}  to have the form 
\begin{eqnarray}
	{{\mathfrak g}^{\sigma\rho}} \rightarrow e^{\pounds_{\xi}}   {\mathfrak g}^{\sigma\rho},
u \rightarrow e^{\pounds_{\xi}}   u,
	\eta_{\mu\nu} \rightarrow \eta_{\mu\nu}
\end{eqnarray}
in terms of the convenient variable ${\mathfrak g}^{\sigma\rho} = \sqrt{-g}  g^{\sigma\rho},$ the flat metric tensor, and matter fields $u$ described by some tensor densities (with indices suppressed).   They made use of a first-order action.  Using a second-order form of the action, we will now confirm that this transformation indeed changes the action merely be a boundary term.  We will also derive convenient formulas involving different sets of variables.    In addition, we will introduce a formula for finite gauge transformations of tetrad fields.

	We recall the bimetric form of the action above for a generally covariant theory\cite{SliBimGRG}, with the metric here expressed in terms of the weight 1 inverse metric: 
\begin{equation}
S = S_{1} [{\mathfrak g}^{\mu\nu}, u] + \frac{1}{2} \int d^{4}x R_{\mu\nu\rho\sigma} (\eta)
{\mathcal{M}} ^{\mu\nu\rho\sigma} + 2 b \int d^{4}x \sqrt{-\eta}  + \int d^{4}x \partial_{\mu}
\alpha^{\mu}.
\end{equation}
Clearly the terms other than $S_{1}$ change at most by a boundary term, so our attention turns to  $S_{1} = \int d^{4}x {\mathcal{L}} _{1} $. 
 The important term $ {\mathcal{L}} _{1}$ in the Lagrangian density is 
just the sum of terms which are products of 
${\mathfrak g}^{\mu\nu}$, $u$, and their partial derivatives.

We now derive a useful formula.  Writing out $e^{\pounds_{\xi}}   A$  as a series $e^{\pounds_{\xi}}   A = \sum_{i = 0}^{\infty} \frac{1}{i!} \pounds_{\xi}^{i} A$
for some tensor density $A$ will put us in a position to derive a useful `product' rule for the exponential of Lie differentiation.  One could write a similar series for another tensor density $B$.  Multiplying the series and using the Cauchy product formula \cite{ChurchillComplex}
\begin{equation}
\sum_{i = 0}^{\infty} a_{i} z^{i} \sum_{j = 0}^{\infty} b_{j} z^{j} = 
\sum_{n = 0}^{\infty}   \sum_{k = 0}^{n} a_{k} b_{n-k} z^{k} 
\end{equation} and the $n$-fold iterated Leibniz rule \cite{ChurchillComplex}
\begin{equation}
[fg]^{(n)} =    \sum_{k = 0}^{n} \frac{n!}{k!(n-k)!} f^{(k)} g^{(n-k)},
\end{equation} 
one recognizes the result as the series expansion of $e^{\pounds_{\xi}}  (AB)$, so one has the pleasant result 
\begin{equation}
(e^{\pounds_{\xi}} A)   (e^{\pounds_{\xi}} B) = e^{\pounds_{\xi}} (AB)
\end{equation}

 Using the fact that partial differentiation commutes with Lie differentiation, 
 ones sees that replacing
 ${\mathfrak g}^{\mu\nu}$ by $e^{\pounds_{\xi}} {\mathfrak g}^{\mu\nu}$ and $u$ with  $e^{\pounds_{\xi}} u$ in ${\mathcal{L}} _{1}$ will give $e^{\pounds_{\xi}} {\mathcal{L}}_{1}$.  Thus, the change in $ {\mathcal{L}} _{1}$  is $\delta {\mathcal{L}} _{1} = (e^{\pounds_{\xi}} -1) {\mathcal{L}} _{1}$,
 which is the Lie derivative of a scalar density of weight 1.  Recalling \cite{SliBimGRG} that the Lie derivative  a weight 1 scalar density $\phi$ is  $\pounds_{\xi} \phi = ( \xi^{\mu} \phi ),_{\mu}$, one sees that $\delta {\mathcal{L}} _{1} $ is just a coordinate divergence, as desired.

	In view of the matrix relationships among the various metric quantities, one has by definition that $( {\mathfrak g}^{\mu\nu} + \delta {\mathfrak g}^{\mu\nu}) ( {\mathfrak g}_{\rho\nu} + \delta {\mathfrak g}_{\rho\nu})  = \delta^{\mu}_{\rho}$ and various other relations.  In that way, one can derive the form of 
 $\delta {\mathfrak g}_{\rho\nu} $, 
  $\delta g$,  $ \delta g_{\rho\nu} $, and the like.  Let us show this fact explicitly for $ g$,   using ${\mathfrak g}^{\sigma\rho} + \delta {\mathfrak g}^{\sigma\rho} =   e^{\pounds_{\xi}}   {\mathfrak g}^{\sigma\rho}.$
The determinant is given by $| {\mathfrak g}^{\sigma\rho}| = [\alpha\mu\nu\rho] \delta^{0}_{\beta} \delta^{1}_{\chi} \delta^{2}_{\psi} \delta^{3}_{\phi} {\mathfrak g}^{\alpha\beta} {\mathfrak g}^{\mu\chi} {\mathfrak g}^{\nu\psi} {\mathfrak g}^{\rho\phi}, $ where $[\alpha\mu\nu\rho] $ is the totally antisymmetric symbol with $[0123] =1.$   Because this form for the determinant holds in any coordinate system, $ [\alpha\mu\nu\rho] \delta^{0}_{\beta} \delta^{1}_{\chi} \delta^{2}_{\psi} \delta^{3}_{\phi} $ is a \emph{scalar} (and also a constant), so  $    e^{\pounds_{\xi}}    ( [\alpha\mu\nu\rho] \delta^{0}_{\beta} \delta^{1}_{\chi} \delta^{2}_{\psi} \delta^{3}_{\phi}  )       = [\alpha\mu\nu\rho] \delta^{0}_{\beta} \delta^{1}_{\chi} \delta^{2}_{\psi} \delta^{3}_{\phi} .$
We therefore have 
\begin{eqnarray}
|   e^{\pounds_{\xi}}        {\mathfrak g}^{\sigma\rho}| = [\alpha\mu\nu\rho] \delta^{0}_{\beta} \delta^{1}_{\chi} \delta^{2}_{\psi} \delta^{3}_{\phi} ({ e^{\pounds_{\xi}} \mathfrak g}^{\alpha\beta}) ( { e^{\pounds_{\xi}}   \mathfrak g}^{\mu\chi}) ( { e^{\pounds_{\xi}}  \mathfrak g}^{\nu\psi}) ( {  e^{\pounds_{\xi}}      \mathfrak g}^{\rho\phi}) \nonumber \\
=  e^{\pounds_{\xi}}      |     {\mathfrak g}^{\sigma\rho}| .
\end{eqnarray}
Using $ {\mathfrak g}^{\sigma\rho} =  g^{\sigma\rho} \sqrt{-g},$ one recalls that $| {\mathfrak g}^{\sigma\rho} | =     | g_{\sigma\rho} | , $
so
\begin{equation}
g + \delta g = e^{\pounds_{\xi}}      g.
\end{equation}
The relation $ -g - \delta g = (\sqrt{-g} + \delta \sqrt{-g} )^{2}$ defines $\delta \sqrt{-g}, $ so one quickly also finds that  
\begin{equation}
\sqrt{-g} + \delta \sqrt{-g} =  e^{\pounds_{\xi}}      \sqrt{-g} ,
\end{equation}
with which one readily finds the result for $ g^{\sigma\rho} $ and so on.  
 Again the transformed field is just the exponentiated Lie derivative of the original.  One therefore can readily use variables other than the densitized inverse metric.  

	Grishchuk, Petrov, and Popova  
have exhibited a straightforward and attractive relationship between finite gauge transformations (with the exponentiated Lie differentiation) and the tensor transformation law \cite{GrishchukPetrov,PopovaPetrov}.  Evidently the former is fundamental, the latter derived. 
 One can define a vector field $\xi^{\alpha}$ using the fact that under a gauge transformation, ${\mathfrak{g}}^{\mu\nu}$ changes in accord with the tensor transformation law, while the flat metric stays fixed.  Let us follow them and define $\xi^{\alpha}$  in terms of the finite coordinate transformation
\begin{equation}
 x^{\prime \alpha} = e^{  \xi^{\mu} \frac{\partial}{\partial x^{\mu} } } x^{\alpha}.
\end{equation}
Then the tensor transformation law, which is easy to use, gives the finite gauge transformation formula, which is difficult to use.  One can therefore hope to avoid using the latter at all, and work only with the tensor 
transformation law, which will tend to be simpler  and will relate to known results in the geometrical theory.  

	It appears that finite gauge transformations for an orthornormal tetrad formalism have never been studied before, so let us do so.  If one imposes no requirements on the tetrad other than that it be orthonormal, then the formula is nonunique in the local Lorentz transformation matrix.  The desired form turns out to be 
\begin{equation}
e^{\mu}_{A} + \delta e^{\mu}_{A} = e^{\pounds_{\xi} } (e^{F})^{C}_{A} e^{\mu}_{C}.
\end{equation}
The Lie differentiation in the first factor acts on everything to its right.  
$F$ is a matrix field which, when an index is moved using $\eta_{AB} = diag(-1,1,1,1)$  or $\eta^{AB},$ is antisymmetric: $F_{A}^{C}= -\eta_{AE} F^{E}_{B}\eta^{BC}.$ 
It is not difficult to verify that the above formula preserves both the completeness
 relation to the inverse metric 
$g^{\mu\nu} = e^{\mu}_{A} \eta^{AB} e^{\nu}_{B}$ and the orthonormality relation $g_{\mu\nu} e^{\mu}_{A} e^{\nu}_{B} = \eta_{AB}.$ Let us now verify the completeness relation by showing that this relation with the gauge-transformed tetrad yields the gauge-transformed curved metric, using $(e^{\pounds_{\xi}} A)   (e^{\pounds_{\xi}} B) = e^{\pounds_{\xi}} (AB).$ One has by definition of a variation $\Delta$ induced by this tetrad transformation,
\begin{eqnarray}
g^{\mu\nu} + \Delta g^{\mu\nu} = (e^{\mu}_{A} + \delta e^{\mu}_{A})  \eta^{AB} (e^{\nu}_{B} + \delta e^{\nu}_{B}) 
\nonumber \\
 = [e^{\pounds_{\xi} } (e^{ F} )^{C}_{A} e^{\mu}_{C}]	 	\eta^{AB}		 e^{\pounds_{\xi} } (e^{ F})^{E}_{B} e^{\nu}_{E}  \nonumber \\
= e^{\pounds_{\xi} } [(e^{ F})^{C}_{A} e^{\mu}_{C}	 	\eta^{AB}		  (e^{ F})^{E}_{B} e^{\nu}_{E}  ].
\end{eqnarray}
Acting with $e^{-\pounds_{\xi} } $ gives
\begin{eqnarray}
 e^{-\pounds_{\xi} }    (g^{\mu\nu} + \Delta g^{\mu\nu} ) = (e^{ F })^{C}_{A} e^{\mu}_{C}	 	\eta^{AB}		  (e^{ F })^{E}_{B} e^{\nu}_{E}.
\end{eqnarray}
  We shall use the  near-antisymmetry of $F$:  $F_{E}^{C} = - \eta_{EJ} F^{J}_{B} \eta^{BC}.$  One then has
\begin{eqnarray}
 e^{-\pounds_{\xi} }    (g^{\mu\nu} + \Delta g^{\mu\nu} )  =   e^{\mu}_{C} (e^{F})^{C}_{A} \eta^{AB} (e^{F})^{E}_{B}    e^{\nu}_{E}  \nonumber \\
=   e^{\mu}_{C} (e^{F})^{C}_{A} \eta^{AB} (I^{E}_{B}  + F^{E}_{B}  + F^{E}_{J} F^{J}_{B} + \ldots  )  e^{\nu}_{E} , 
\end{eqnarray}
where the one factor is expanded as a series.  Continuing by moving the Lorentz matrix into strategic locations gives 
\begin{eqnarray}   e^{\mu}_{C} (e^{F})^{C}_{A}  (I^{A}_{P} + \eta^{AB} F^{E}_{B} \eta_{EP} +  \eta^{AB}  F^{J}_{B} \eta_{JK} \eta^{KL} F^{E}_{L} \eta_{EP} + \ldots ) e^{P\nu} \nonumber \\
=  
 e^{\mu}_{C} (e^{F})^{C}_{A}  (I^{A}_{J} - F^{A}_{J}  +   F^{A}_{K} F^{K}_{J} - \ldots ) e^{J\nu}, 
\end{eqnarray}
where the near-antisymmetry of $F$ has been employed.
Reverting to the exponential form gives
\begin{eqnarray} 
 e^{\mu}_{C} (e^{F})^{C}_{A} (e^{-F})^{A}_{J} e^{J\nu} \nonumber \\
=  
 e^{\mu}_{C} I^{C}_{E} e^{E\nu}  \nonumber \\
= g^{\mu\nu},
\end{eqnarray}
leading to the expected conclusion  $g^{\mu\nu} + \Delta g^{\mu\nu}  =   g^{\mu\nu} + \delta g^{\mu\nu}  = e^{\pounds_{\xi} }    g^{\mu\nu}.$  Thus, completeness holds, and the tetrad-induced variation $\Delta$ of the inverse curved metric agrees with the gauge transformation variation $\delta.$  
By similar maneuvers, one establishes the orthonormality relation for this tetrad variation:
\begin{eqnarray} 
 (  e^{\pounds_{\xi} }    g_{\mu\nu} )  (e^{\mu}_{A} + \delta e^{\mu}_{A}) (e^{\nu}_{B} + \delta e^{\nu}_{B}) =  \eta^{AB}.
\end{eqnarray}
Finally, the inverse tetrad transforms as 
\begin{equation}
f^{A}_{\mu} + \delta f^{A}_{\mu} = e^{\pounds_{\xi} } (e^{  -F })^{A}_{C} f^{C}_{\mu},
\end{equation}
with a negative sign applied to $F$.

 \subsection{Gauge Transformations Not a Group, But a Groupoid}

	If one is not interested in taking $\eta$-causality seriously, then any suitably smooth vector field, perhaps subject to some boundary conditions, will generate a gauge transformation.    This is the notion that Grishchuk, Petrov, and Popova have employed, and that on occasion we have used above.  
  However, in the SRA, respecting $\eta$-causality--indeed, preferably stable $\eta$-causality--is essential.  This fact entails that only a subset of all vector fields generates gauge transformations in the SRA. 

	Let us be more precise in defining gauge transformations in the SRA, requiring stable $\eta$-causality instead. A gauge transformation in the SRA is a mathematical transformation generated by a vector field in the form 
\[
g_{\mu \nu }\rightarrow e^{\pounds _{\xi}}g_{\mu\nu},
\eta _{\mu \nu}\rightarrow \eta _{\mu\nu},
u\rightarrow e^{\pounds _{\xi }}u, 
\]
but we now introduce the requirement that both the original and the transformed curved metrics 
respect stable $\eta$-causality.  It is evident that a vector field that generates a gauge transformation given one curved metric and a flat metric, might not generate a gauge transformation given another curved metric (and the same flat metric), because in the second case, the transformation might move the curved metric out of the  $\eta$-causality-respecting configuration space, which
 is only a subset of the naive configuration space. 

	It follows that one cannot identify gauge transformations with generating vector fields alone.  Rather, one must also specify the curved metric prior to the transformation.  For thoroughness, one can also use the flat metric as a label, to ensure that the trivial coordinate freedom is not confused with the physically significant gauge freedom.  Let us therefore  write a gauge transformation as an ordered triple involving a vector field, a flat metric tensor field, and a curved metric tensor field:
\begin{equation}
( e^{\pounds_{\xi} }, \eta_{\mu\nu}, g_{\mu\nu}),
\end{equation}
where both $g_{\mu\nu}$ and $e^{\pounds_{\xi} } g_{\mu\nu}$ satisfy stable causality with respect  to $\eta_{\mu\nu}.$ The former restriction limits the configuration space for the curved metric, whereas the latter restricts the vector field. (At this point we drop the indices  for brevity.)  The non-Abelian nature of these transformations makes it advisable to use not $\xi$, but the operator $e^{\pounds_{\xi}}$, to label the transformations, because then the noncommutativity of two transformations is manifest.   

	One wants to compose two gauge transformations to get a third gauge transformation.  At this point, the fact that a gauge transformation is not labelled merely by the vector field, but also by the curved and flat metrics, has important consequences.  Clearly the two gauge transformations to be composed must have the second one start with the curved metric with which the first one stops. 
 We also want the flat metrics to agree.  Thus, the `group' multiplication operation is defined only in certain cases, meaning the gauge transformations in the SRA \emph{do not form a group}, despite the inheritance of the mathematical form of exponentiating the Lie differentiation operator from the field formulation's gauge transformation. 
 Two gauge transformations $(e^{\pounds_{\psi} }, \eta_{2}, g_{2})$ and $(e^{\pounds_{\xi} }, \eta_{1}, g_{1})$ can be composed to give a new gauge transformation
 $(e^{\pounds_{\psi} }, \eta_{2}, g_{2}) \circ (e^{\pounds_{\xi} }, \eta_{1}, g_{1})$  only if $g_{2} = e^{\pounds_{\xi} } g_{1}$ and $\eta_{2} = \eta_{1}.$ 
  The left inverse of $(e^{\pounds_{\xi} }, \eta_{1},  g_{1} )$ is    $(e^{\pounds_{-\xi} }, \eta_{1}, e^{\pounds_{\xi} } g_{1} ),$ yielding
   $(e^{\pounds_{-\xi} }, \eta_{1}, e^{\pounds_{\xi} } g_{1} ) \circ   (e^{\pounds_{\xi} }, \eta_{1},  g_{1} ) = ( 1, \eta_{1}, g_{1}),$ an identity transformation.  The right inverse is  also $(e^{\pounds_{-\xi} }, \eta_{1}, e^{\pounds_{\xi} } g_{1} ),$ yielding 
$ (e^{\pounds_{\xi} }, \eta_{1},  g_{1} ) \circ (e^{\pounds_{-\xi} }, \eta_{1}, e^{\pounds_{\xi} } g_{1} )  = (1, \eta_{1}, e^{\pounds_{\xi} }  g_{1} ),$ which is also an identity  transformation.  Gauge transformations in the SRA do not form a group, because multiplication between some elements simply is not defined.  While the lack of a group structure is perhaps unfamiliar, there is a known mathematical structure that precisely corresponds to what the physics of the SRA dictates.   According to A. Ramsay, ``[a] groupoid is, roughly speaking, a set with a not everywhere defined binary operation, which would be a group if the operation were defined everywhere.'' \cite{Ramsay} (pp. 254, 255)  One need not rest with informal descriptions, because one can easily show that SRA gauge transformations in fact satisfy the axioms required of a groupoid, as defined by P. Hahn \cite{Hahn} and J. Renault \cite{Renault}.  Though groupoids are increasingly coming to the attention of physicists, they are sufficiently obscure that we reproduce the definition of Hahn and Renault here for convenience.  According to them \cite{Hahn,Renault}, a groupoid is a set $G$ endowed with a product map $(x,y)\rightarrow
xy:G^{2}\rightarrow G$, where $G^{2}$ is a subset of $G\times G$ called the
set of composable [ordered] pairs, and an inverse map $x\rightarrow
x^{-1}:G\rightarrow G$ such that the following relations are satisfied:
\begin{enumerate}
\item  $(x^{-1})^{-1}=x$,

\item  if $(x,y)$ and $(y,z)$ are elements of $G^{2}$, then $(xy,z)$ and $(x,yz)$ are elements of $G^{2}$ and $(xy)z=x(yz)$,

\item  $(x^{-1},x)\in G^{2}$, and if $(x,y)\in G^{2}$, then $x^{-1}(xy)=y$,

\item  $(x,x^{-1})\in G^{2}$, and if $(z,x)\in G^{2}$, then $(zx)x^{-1}=z$.
\end{enumerate}
One readily interprets this definition as implying that every SRA gauge transformation has an inverse, and that multiplication is associative whenever it is defined.

	As we saw above, authors such as Grishchuk \cite{Grishchuk90} and A. N. Petrov \cite{Petrov} have denied that the flat metric's null cone can have any physical significance in part  because the relation between the two metrics' null cones is gauge-variant.  While this objection holds if one insists that gauge transformations must form a group, such insistence seems unwarranted.  In accord with the demands of the SRA,  we define gauge transformations such that they respect the causal structure of the flat metric, and find that gauge transformations form a groupoid.   Thus this objection to ascribing physical significance to the flat metric's null cone is removed.

 \subsection{Canonical Quantization in the SRA}

	The relevance of the special relativistic approach to Einstein's equations to canonical quantum gravity deserves some consideration.  The primary patrons of the flat metric in the context of Einstein's equations have been the particle physicists in the context of the old covariant perturbation program of quantization.  However, this program famously appears to be perturbatively nonrenormalizable, even  with the 
addition of carefully chosen matter fields in the later supergravity era \cite{IshamPrima}.  Therefore, the covariant perturbation program has largely been abandoned.\footnote{An exception is some recent work by G. Scharf and collaborators such as I. Schorn, N. Grillo, and M. Wellmann (for example, \cite{Scharf,ScharfWellmann}).  Their use of ``causal'' methods helps to achieve finite results.  Another approach \cite{Lauscher}, based on an old suggestion by S. Weinberg, is that the theory might be  \emph{nonperturbatively} renormalizable.  Recently O. Lauscher and M. Reuter have argued that this situation likely is realized \cite{Lauscher}.}  

	With the patrons of the flat metric having largely diverted their attention to strings, membranes, and the like, one might form the belief that the use of a flat background metric has nothing further to contribute to 
quantum gravity, and in particular, to canonical quantum gravity.  Isham writes of the null cone issue in the covariant perturbation program:  ``This very non-trivial problem is one of the reasons why the canonical approach to quantum 
gravity has been so popular.'' \cite{IshamPrima} (p. 12)  And again, ``One of the main aspirations of the canonical approach to quantum gravity has always been to build a formalism with no background spatial, or spacetime, metric.'' \cite{IshamPrima} (p. 18) The use of a flat background in canonical 
gravity indeed seems to be rather rare, apart from some work in the field formulation by Grishchuk and Petrov \cite{GrishchukPetrov}, which does not consider the flat metric's null cone.  

	However, it would be a mistake to conclude that the canonical formalism is immune to similar worries, worries which a flat background's null cone structure could address.  Isham continues:  
\begin{quote}
	However, a causal problem arises here [in the canonical approach] too.  For example, in the Wheeler-DeWitt approach, the configuration variable of the system is the Riemannian metric $q_{ab}(x)$ on a three-manifold $\Sigma,$ and the canonical commutation relations invariably include the set 
\begin{equation}
[\hat{q}_{ab}(x), \hat{q}_{cd}(x^{\prime} )] = 0
\end{equation}
for all points $x$ and $x^{\prime} $ in $\Sigma.$  In normal canonical quantum field theory such a relation arises because $\Sigma$ is a space-like subset of spacetime, and hence the fields at $x$ and $x \prime$ should be simultaneously measurable.  But how can such a relation be justified in a theory that has no fixed causal structure?  The problem is rarely mentioned but it means that, in this respect, the canonical approach to quantum gravity is no better than the covariant one.  It is another aspect of the `problem of time' \ldots. \cite{IshamPrima} (p. 12)
\end{quote}	
Evidently introducing a flat metric can help:
\begin{quote} The background metric $\eta$ provides a fixed causal structure with the usual family of Lorentzian inertial frames.  Thus, at this level, there is no problem of time.  The causal structure also allows a notion of microcausality, thereby permitting a conventional type of relativistic quantum field theory \ldots 
	It is clear that many of the \emph{prima facie} issues discussed earlier are resolved in an approach of this type by virtue of its heavy use of background structure. \cite{IshamPrima} (p. 17)
\end{quote}
What then is the difficulty?
\begin{quote}
However, many classical relativists object violently \ldots, not least because the background causal structure cannot generally be identified with the physical one.  Also, one is restricted to a specific background topology, and so a scheme of this type is not well adapted for addressing many of the most interesting questions in quantum gravity: black hole phenomena, quantum cosmology, phase changes \emph{etc.} \cite{IshamPrima} (p. 17)
\end{quote}
However, above we have presented a formalism which plausibly ensures ensure that the physical causal structure is consistent with the background one by construction.  Thus, this first objection is answered.  The second objection is strong only if one already knows that
 gravitation is geometrical at the classical level.  But such a view is hardly mandatory.  
We conclude that it would be interesting to investigate the canonical quantization of Einstein's equations within the special relativistic 
approach, because serious conceptual problems with standard approaches would evidently be resolved.

\section{Global Hyperbolicity, Black Hole Information Loss, and a Well Posed Initial Value Formulation}

	With the requirement of $\eta$-causality imposed--perhaps stable $\eta$-causality using the causality variables--it follows that any ``SRA spacetime'' $(R^{4}, \eta_{\mu\nu}, g_{\mu\nu})$ is globally hyperbolic.  How so?  It follows from $\eta$-causality that the future domain of dependence of an $\eta$-spacelike slice is in fact the whole of $R^{4}.$  But global 
hyperbolicity just is the possession of a Cauchy surface \cite{Wald}, so any $\eta$-causal SRA spacetime  $(R^{4}, \eta_{\mu\nu}, g_{\mu\nu})$  is globally hyperbolic.  Global hyperbolicity  resolves the Hawking black hole information loss paradox 
\cite{EarmanInfo}.  Thus, there is no evolution from a pure state to a mixed state, and so no information is lost.   If global hyperbolicity solves the problem, then so does the SRA.  Evidently the geometrical interpretation of Einstein's equations helps to generate this paradox, so the SRA becomes somewhat more attractive on conceptual grounds.

 	If the nature of black holes in the SRA is altered, one might wonder what becomes of the work on black hole entropy.  As J. Oppenheim has shown recently \cite{Oppenheim}, the proportionality of black hole entropy to area does not depend on the existence of an event horizon, 

but merely occurs in the limit as a gravitating system approaches its gravitational radius.  Inclusion of the gravitational field in thermodynamics yields a correction term that violates entropy extensivity; in the limit as the radius approaches the Schwarzschild radius, the entropy is proportional to area rather than volume.

	One could further ask whether the SRA has a well posed initial value formulation.  
The use of  harmonic coordinates has been a common technique for answering this question in the geometrical formulation \cite{Wald}, in which the choice of harmonic coordinates constitutes a gauge-fixing.  Given that the SRA has somewhat restricted gauge
 freedom, and that the tensorial DeDonder gauge condition (which makes the coordinate $g$-harmonic when $\eta$-Cartesian) has causal difficulties for plane wave solutions, one might fear that the proofs of a well posed initial value formulation fail for the SRA.  However, such a fear is groundless, as we see if we resist the temptation
to use $\eta$-Cartesian coordinates, for which we have no need.  The SRA has both coordinate freedom and gauge freedom.  The choice of $g$-harmonic coordinates, if the flat metric is unfixed, is merely a coordinate choice, leaving the gauge unfixed.  The eigenvalues, which express the relation between the two null cones, are coordinate scalars. The naive gauge freedom has not been used at all, and thus is fully available for deforming the curved metric until it becomes consistent with the flat one by reducing the lapse.  Therefore the traditional harmonic coordinate approach to demonstrating a well posed initial value problem experiences no obstacles from the SRA. The SRA indeed has a well posed initial value formulation, and one need not even appeal to recent work that permits coordinate freedom \cite{ACBYork} to show it.


\section{Conclusion}

	We have aimed to take special relativity seriously, including its causal structure, while viewing gravity as described by Einstein's field equations.  In  reviewing the history of the Lorentz-covariant approach, we found that the fundamental issue of causality has frequently been ignored or distorted, and evidently has never been handled correctly.  Next we found an adequate kinematical description of the relation between the null cones, and found that all metrics obeying stable $\eta$-causality, and almost all metrics obeying $\eta$-causality, possess in effect an orthonormal tetrad of eigenvectors.  Plausibly one can deforming any physically relevant solution into one in which the proper null cone relation obtains.  Having done so, one can adopt a new set of variables which ensure that the proper relation holds automatically.  Gauge transformations form not a group, but a groupoid.  As a result of using the flat metric, the problem of defining causality in quantum gravity is solved.  Furthermore, every SRA solution is globally hyperbolic, so the Hawking black hole information loss paradox does not to arise.



\end{document}